# Optimization of Executable Formal Interpreters developed in Higher-order Theorem Proving Systems

Zheng Yang[1*] Hang Lei
zyang.uestc@gmail.com hlei@uestc.edu.cn

[1]School of Information and Software Engineering, University of Electronic Science and Technology of China,
No.4, Section 2, North Jianshe Road, 610054, Sichuan, Chengdu, P.R. China.

**Abstract**. In recent publications, we presented a novel formal symbolic process virtual machine (FSPVM) framework that combined higher-order theorem proving and symbolic execution for verifying the reliability and security of smart contracts developed in the Ethereum blockchain system without suffering the standard issues surrounding reusability, consistency, and automation. A specific FSPVM, denoted as FSPVM-E, was developed in Coq based on a general, extensible, and reusable formal memory (GERM) framework, an extensible and universal formal intermediate programming language, denoted as Lolisa, which is a large subset of the Solidity programming language that uses generalized algebraic datatypes, and a corresponding formally verified interpreter for Lolisa, denoted as FEther, which serves as a crucial component of FSPVM-E. However, our past work has demonstrated that the execution efficiency of the standard development of FEther is extremely low. As a result, FSPVM-E fails to achieve its expected verification effect. The present work addresses this issue by first identifying three root causes of the low execution efficiency of formal interpreters. We then build abstract models of these causes, and present respective optimization schemes for rectifying the identified conditions. Finally, we apply these optimization schemes to FEther, and demonstrate that its execution efficiency has been improved significantly.

## 1. Introduction

Blockchain technology [1], such as the Ethereum blockchain system, has been adopted in a wide variety of applications such as cryptocurrency [2] and distributed storage [3]. Among the most widely adopted blockchain systems, Ethereum implements a general-purpose Turing-complete programming language denoted as Solidity [4]. Solidity allows for the development of arbitrary applications and scripts (i.e., programs) that are collectively denoted as smart contracts, which can be executed in a virtual runtime environment denoted as the Ethereum Virtual Machine (EVM) to conduct blockchain transactions automatically. In addition to smart contracts, a number of other lightweight programs have been recently deployed in critical domains. The growing use of these lightweight programs has led to increased scrutiny of their security because they include properties ranging from transaction-ordering dependencies to mishandled exceptions that make them susceptible to deliberate attacks that can result in direct economic loss [5]–[7]. Therefore, it is crucial to verify the security and reliability of such programs in the most rigorous manner available. Among the available verification approaches, higher-order logic theorem proving is one of the most rigorous technologies for verifying the properties of programs. This approach involves establishing a formal model of a software system, and then verifying the system according to a mathematical proof of the formal model. However, this process suffers from problems associated with consistency, reusability, and automation. One of the available solutions for addressing these problems involves designing a formal symbolic process virtual machine (FSPVM) based on higher-order theorem proving technology.

The design and building of a general and powerful FSPVM for certifying and verifying smart contracts operating on multiple blockchain platforms has been an ongoing project undertaken by the present authors for some time. In our recent work [8], we presented a theoretical FSPVM framework based on our proposed extension of Curry-Howard isomorphism, denoted as execution-verification isomorphism (EVI), for automatically verifying lightweight programs and solving the issues associated with reusability, consistency, and automation in higher-order theorem proving technology. Specifically, the proposed theoretical FSPVM framework contains four key elements: EVI, a formal general memory model, a high-level formal intermediate language that is equivalent to high level programming languages in the real world, and a respective formally verified definitional interpreter. Subsequently, we adopted the proposed theoretical FSPVM framework to build an FSPVM, denoted as FSPVM-E, in Coq for the verification of Ethereum smart contracts [9]. FSPVM-E was constructed using a general, extensible, and reusable formal memory (GERM) framework, an extensible intermediate programming language denoted as Lolisa [10], which is a large subset of the Solidity programming language with a mechanized syntax and semantics, and a respective formal interpreter denoted as FEther. The FEther interpreter is the final critical component of FSPVM-E that integrates the trusted core of Coq, GERM, and Lolisa. However, our past work has demonstrated that, if FEther is designed according to the standard approach recommended by most relevant research studies and tutorials regarding programming language formalism and interpreter design (e.g., [22]), its symbolic execution efficiency is extremely low. This low execution efficiency of FEther directly influences the verification efficiency and the automation level of FSPVM-E because FEther is employed for parsing the domain specific language (i.e., Lolisa in the present development), implementing program behavior, modifying the formal memory space, and generating the final logic memory state for program verification, and therefore serves as the proof engine of the overall FSPVM-E framework. As such, this is a crucial issue that must be addressed.

The present work addresses this crucial issue associated with FEther by building a general abstract formal interpreter model to analyze and optimize the design of formal interpreters built in higher-order logic theorem proving assistants. Analysis identifies three essential causes of the low execution efficiency of FEther, which are denoted as *call-by-name termination* (CBNT), *information redundancy explosion* (IRE), and *concurrent reduction* (CR). Next, we build abstract models specific to CBNT, IRE, and CR, and analyze the models in detail to provide respective methods for addressing each of these issues. Finally, we apply these schemes to optimize FEther, and demonstrate that the execution and verification efficiency of FSPVM-E are improved significantly.

The remainder of this paper is structured as follows. Section 2 introduces relevant past studies regarding the formal verification of virtual machines and programs. Because the nature of our present work regarding the optimization of FEther requires that we first introduce some essential concepts and definitions introduced in our past work, Section 3 elaborates on the foundational concepts and definitions required by the present work. Section 4 presents the respective abstract models and analyses specific to CBNT, IRE, and CR. Section 5 describes the solutions established for each issue, and presents experimental verification results based on example smart contracts obtained using FEther after optimization. Section 6 presents the conclusions of our work and the directions of our future efforts.

## 2. Related Work

The work of this paper was primarily inspired by the symbolic process virtual machine KLEE [11], which is a well-known and successful certification tool based on symbolic execution. However, it must be noted that many recent tools are based on symbolic execution [12], but most of them adopt model checking technology as their foundation, and few are developed in a higher-order logic theorem proving system to enable real-world programs to be symbolically executed, and their properties verified automatically using the execution result. However, FSVPM-E supports the higher-order verification of complex properties. In addition, while the verification efficiency of these previously developed symbolic execution tools is reasonably high, we can expect that FSVPM-E will provide a similarly high verification efficiency when properly optimized. Moreover, the fundamental theory of the FEther is a type of higher-order predicate logic which can inductively express all execution situations of a program. Compared with the traditional testing technologies, the FEther satisfied the completeness.

In addition to general verification tools, a number of well-known projects have been developed for the verification of Ethereum smart contracts. For example, the formal semantics denoted as KEVM [16] were developed for the formal low-level programming verification of Solidity bytecode on the EVM platform using the K-framework, like the formalization conducted in Lem [17]. Because KEVM is executable, it can run the validation test suite provided by the Ethereum foundation. However, the low-level verification conducted by KEVM makes it poorly suited to high-level programming languages, such as Solidity, which was a primary motivation for the development of FSVPM-E.

A number of interesting projects have been undertaken that employ higher-order theorem proving assistants as the fundamental platform. For example, Frama-C [18] is an extensible and collaborative platform dedicated to the source-code analysis of software written in the C programming language. In addition, VST [19] is one of the powerful program verification toolchains based on the CompCert project [20], and SMTCoq [21] is another interesting project for developing automatic theorem proving tools. Unfortunately, these platforms also fail to provide a suitable combination of symbolic execution and higher-order theorem proving. Moreover, these studies fail to discuss the verification efficiencies obtained and the optimization schemes employed in their work.

In light of the above analysis of past studies, we note that the present work represents the first systematic discussion regarding the optimization of symbolic execution efficiency in higher-order theorem proving assistants, such as Coq and Isabelle. Moreover, the effect of these optimization schemes is also confirmed by their application to a formally verified interpreter.

## 3. Foundational concepts and definitions

Real world virtual machines of high-level programming languages, such as Smalltalk, Java, and. Net, typically support bytecode as their instruction set architecture (ISA) and are implemented by translating the bytecode for commonly used code paths into native machine code. In contrast, an FSPVM takes the specification functional programming language (*FPL*) provided by a higher-order theorem proving assistant, such as Gallina in Coq, as the bytecode, the formal memory framework (*FMemory*) as the memory space, and the trusted core of the proving assistant as the CPU. However, the trusted core of a proof assistant has only two functions, i.e., evaluating and proving. As such, the fundamental environment provided by Coq is not sufficient to symbolically execute programs written by a mainstream higher-order programming language $\mathcal{L}$, and thereby obtain logic memory states. Therefore, the proof environment of higher-order theorem proving assistants must be extended. A blueprint for the previously proposed logic memory state generation process [8] is illustrated in Figure 1, where a high-level formal intermediate language $\mathcal{FL}$, which is equivalent to $\mathcal{L}$, is adopted for rewriting a real-world program (*RWprogram*) as a formal *RWprogram* (*FRWprogram*), and the respective formally verified interpreter ($\mathcal{FI}$) is formalized in the FSPVM. The executable semantics of $\mathcal{FL}$ play the role of the ISA of $\mathcal{FI}$. In addition, $\mathcal{FI}$ plays the role of the core of the execution engine in the FSPVM whose task is to simulate the real execution process of *FRWprograms* and generate logic memory states. FSPVM-E is wholly developed in Coq with *FMemory* based on the GERM model, $\mathcal{FL}$ specified as Lolisa, and $\mathcal{FI}$ specified as FEther.

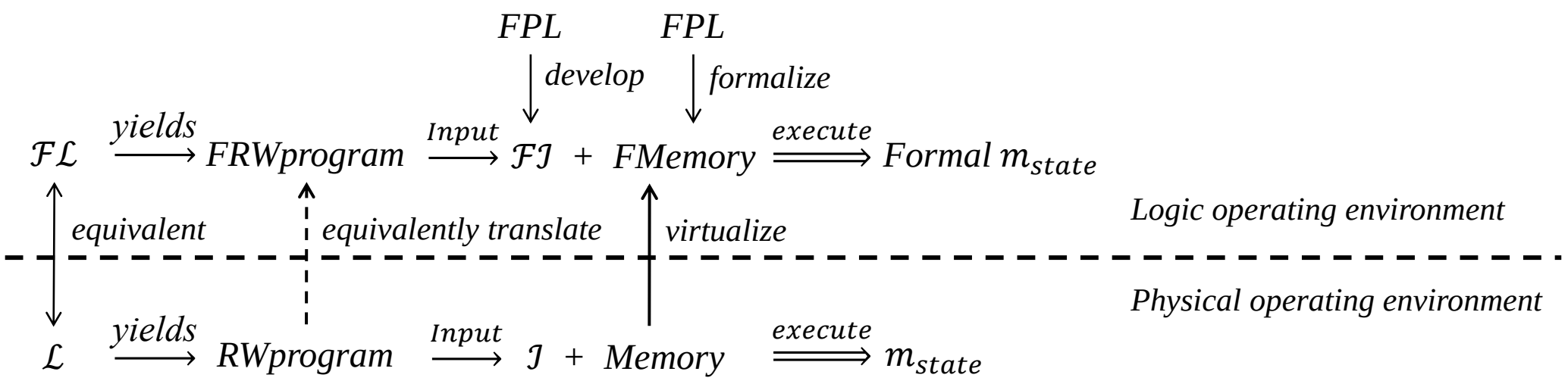

Figure 1. Equivalence between the execution of a real world program (*RWprogram*) written in a higher-order programming language $\mathcal{L}$ and execution in a logic environment using a high-level formal intermediate language $\mathcal{FL}$, which is equivalent to $\mathcal{L}$, to rewrite *RWprogram* as a formal *RWprogram* (*FRWprogram*) in conjunction with a formal memory framework (*FMemory*) as the memory space and a respective formally verified interpreter ($\mathcal{FI}$)

The abstract syntax of Lolisa includes contract declaration (*Contract*), modifier declaration (*Modifier*), variable declaration (*Var*), structure declaration (*Struct*), assignment (*Assign*), return (*Return*), multi-value return (*Returns*), throw (*Throw*), skip (*Snil*), function definition (*Fun*), while loop ($Loop_{while}$), for loop ($Loop_{for}$), function call ($Fun_{call}$), conditional (*If*), and sequence (*Seq*) statements. However, the issues specific to CBNT, IRE, and CR, which form the basis of the present work, are exclusively related to only *Seq* statements. Therefore, only the syntax details of the *Seq* constructor are explicitly defined in Figure 2. Details regarding the other statements employed by Lolisa are reported in our previous work [10]. In addition, the development of FEther in Coq and its verification process will be simplified if the *FRWprograms* written in Lolisa are maintained as structural programs. To ensure this condition, the semantics of Lolisa are made to adhere to the following pointer counter axiom.

**Axiom** (*Pointer Counter*) Suppose that, for all statements *s*, if *s* is the next execution statement, it must be the head of the statement sequence in the next execution iteration.

| |
|---|
| *Statement:* $s ::=$ *Contract* \| *Modifier* \| *Var* \| *Struct* \| *Assignv* |
| \| *Return* \| *Returns* \| *Throw* \| *Snil* |
| \| *Fun* \| $Loop_{while}$ \| $Loop_{for}$ \| $Fun_{call}$ \| *If* \| $Seq(s,\ s')$ |

Figure. 2. Abstract syntax of Lolisa sequence (*Seq*) statements

Table 1 summarizes the helper functions used in the dynamic semantic definitions. Table 2 lists the state functions used to calculate commonly needed values from the current state of programs. All of these state functions will be encountered in the following discussion. Components of specific states will be denoted using the appropriate Greek letter subscripted by the state of interest. As shown in Table 2, the context of the formal memory space is denoted as $M$, where $\sigma$ is employed to denote a specific memory state, and the context of the execution environment is represented as $\mathcal{E}$. Furthermore, we assign $\Omega$ as the native value set of the basic logic system. Also, the proof evaluation will execute in the proof contexts, which we will denote as $\Gamma$, $\Gamma_1$, etc. For brevity in the following discussion, we will assign $\mathcal{F}$ to represent the overall formal system. All the following subsections present semantics evaluation relations of the form $\sigma_0 \Downarrow_{stt} \langle\sigma_1\rangle$, where $\sigma_0$ and $\sigma_1$ are the initial and final memory states, respectively, $v_0$ represents the form of Lolisa syntax being defined, and the nature of $v_1$ depends on the precise evaluation relation being defined. The terms $env$ and $fenv$ represent the current execution environment and the super environment, respectively.

Table 1. State functions

| | | | |
|---|---|---|---|
| $\mathcal{E}$ | environment information | $\mathcal{F}$ | formal system world |
| $M$ | memory space | $\Gamma$ | proof context |

Table 2. Helper functions

| | | | |
|---|---|---|---|
| $set_{env}$ | Changes the current environment | $env_{check}$ | Validates the current environment |
| *Statement outcomes:* *out* $::=$ | *normal* | *continue with the next statement* | |
| | \| *stop* | *stop executing current statement* | |
| | \| *error* | *stop executing current statement with error message* | |
| | \| *exit* | *function exit* | |

## 4. Problem analysis

As discussed, the computational efficiency of the previous development of FEther was not sufficiently high to execute and verify formal programs written in Lolisa. A simple example of this can be illustrated by the conditional statement given as Example 1, as follows:

$$if_{throw} \stackrel{\text{def}}{=} \forall\left(s, s' : statement\right), if\ (true)\{\ throw\ ();\ \}else\ \{s;\}s' \quad \text{, (Example 1)}$$

where $throw\ ()$ is a widely defined special function in programming languages like JavaScript that is called to throw out an executing program. This simple code segment will execute $throw\ ()$ to throw out an executing program and return the initial memory state $m_{init}$. However, as shown in the Figure 3, executing (i.e., verifying) this very simple code segment using the non-optimized development of FEther requires an execution time of 92.546 s, which is unacceptably long.

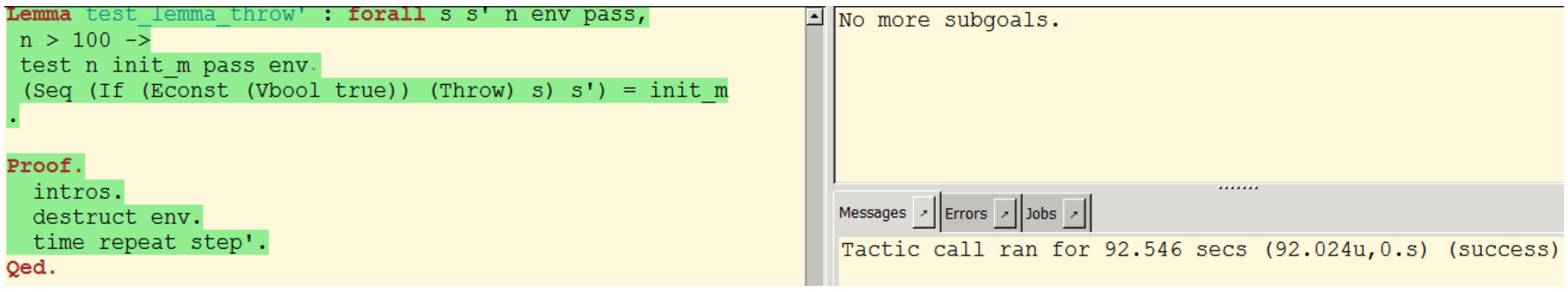

Figure 3. Evaluation time required for Example 1 by the non-optimized FEther

First, we must obtain an objective appraisal of the computational efficiency of FEther. To this end, we employed the example smart contracts given in [4] as the testing data set, and evaluated the maximum execution time required by FEther. These example smart contracts include from 0 to 35 lines both with quantifier abstraction (like Example 1) and without quantifier abstraction. We employed 5 identical personal computers with equivalent hardware of 8 G memory and a 3.20 GHz CPU, and equivalent software, including Windows 10 and CoqIDE 8.6. Each computer executed the same data set 100 times to obtain the peak execution times of the FEther evaluation process. The results are shown in Figure 4. If we set an execution time limitation of 3600 s, programs written in Lolisa without quantifier abstraction that are greater than 30 lines will exceed the limitation, and programs with quantifier abstraction that are greater than about 15 lines will exceed the limitation.

Through analyzing the results given in Figure 4, we can determine that the standard implementation of FEther, which requires that sequence statements be defined explicitly, will generate a volume of logic information in the current context $\Gamma_c$ that will exceed the affordable range of higher-order theorem proving assistants, making the evaluation efficiency extremely low. As discussed, the extremely low computational efficiency is specifically caused and exacerbated by the three crucial problems CBNT, IRE, and CR. These problems are analyzed and defined in the following subsections.

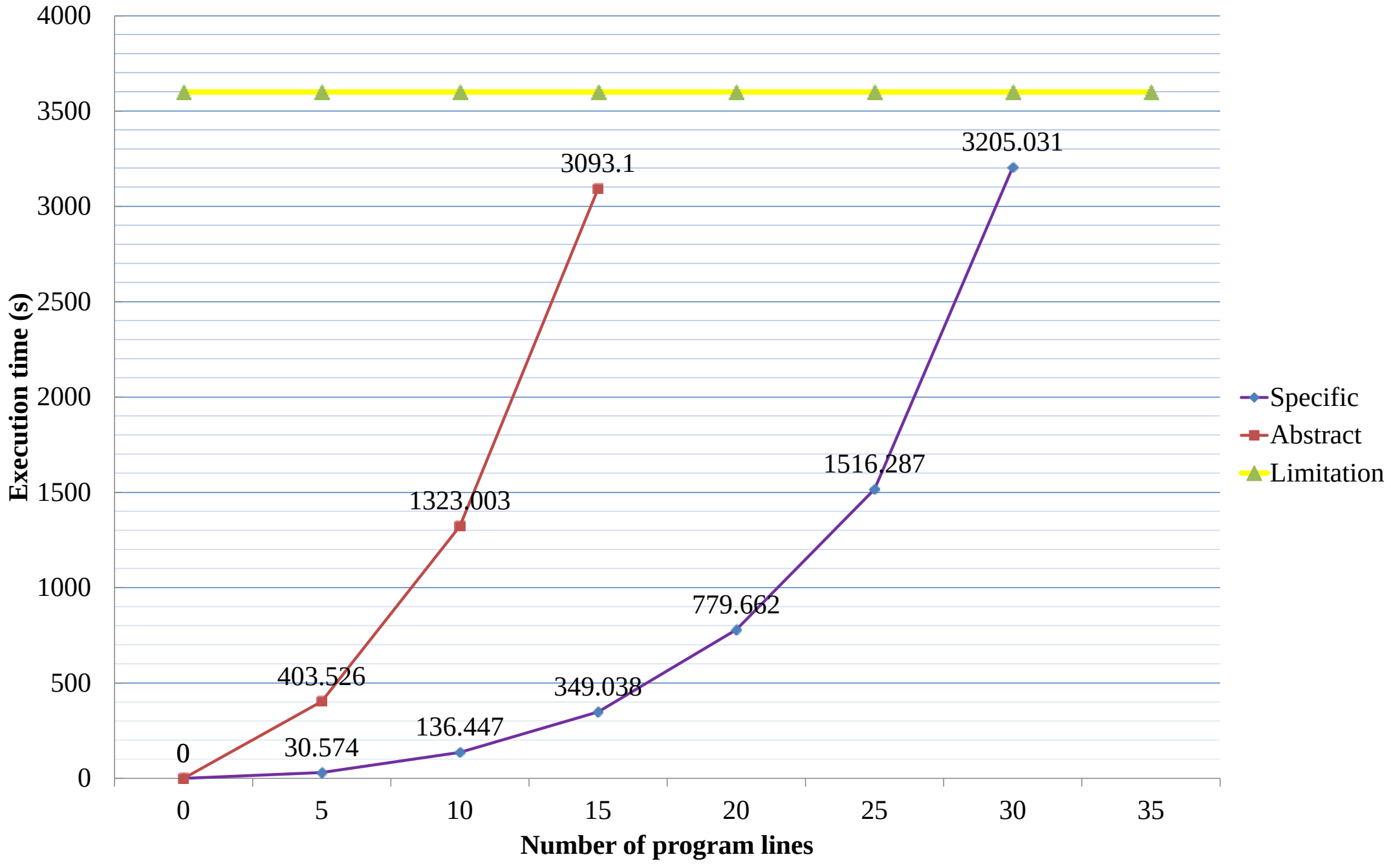

Figure 4. Maximum evaluation times of FEther for example smart contracts given in [4]

***Problem* 1** (***call-by-name termination***) The root of CBNT is caused by the evaluation strategy of lambda calculus while FEther is evaluating the semantics of *Seq* statements.

First, we note that the essence of a formal interpreter $\mathcal{FI}$ employed as part of an FSPVM framework is a large recursive function written in the specification language provided by a higher-order theorem proving system. The type of $\mathcal{FI}$ can be abstractly defined as follows.

$$\mathcal{FI} :: args \rightarrow optional\ memory \rightarrow FRWprogram \rightarrow optional\ memory \quad (1)$$

Therefore, the symbolic execution process $P_{exe}$ is equivalent with the process of evaluating recursive functions $P_{eval}$ in a higher-order theorem proving system, as indicated by the following expression.

$$\Omega, M, \mathcal{F} \vdash_{ins} P_{exe} \equiv P_{eval} \quad (2)$$

Next, we note that the *Seq* statement is one of the most essential statements that is used in formal semantics, including operational semantics, denotational semantics and axiom semantics, to connect the remaining types of statements. In most relevant research studies and tutorials regarding programming language formalism and interpreter design (e.g., [22]), it is standard to explicitly define the abstract syntax and semantics of *Seq* statements using *Seq* constructors. For example, the formal semantics of *Seq* statements in Lolisa are defined explicitly according to rules EVAL-STT-SEQ1 and EVAL-STT-SEQ2 below [10]. In Coq, statements are mechanized as an inductive type *statement* that is constructed by specific statement constructors, representing statement tokens. Therefore, an *FRWprogram* can be summarized according to expression (3) below.

$$\frac{\begin{array}{c} M \vdash \sigma \qquad \mathcal{E}, M, \mathcal{F} \vdash s_0, s_1 \\ s_0 \neq Seq\left(s, s'\right) \wedge \sigma \Downarrow_{s_0} \left(\sigma', normal\right) \\ env_{check}(env, fenv) \hookrightarrow true \wedge set_{env}(Seq(s_0, s_1), env) \hookrightarrow Some\ env' \end{array}}{\mathcal{E}, M, \mathcal{F} \vdash \langle \sigma, env, fenv, Seq(s_0, s_1) \rangle \Rightarrow \langle \sigma' \Downarrow_{s_1}, env', fenv, normal \rangle} \quad \text{(EVAL-STT-SEQ1)}$$

$$\frac{\begin{array}{c} M \vdash \sigma \qquad \mathcal{E}, M, \mathcal{F} \vdash s_0, s_1 \\ s_0 \neq Seq\left(s, s'\right) \wedge \sigma \Downarrow_{s_0} \left(\sigma', error\right) \\ env_{check}(env, fenv) \hookrightarrow true \wedge set_{gas}(Seq(s_0, s_1), env) \hookrightarrow Some\ env' \end{array}}{\mathcal{E}, M, \mathcal{F} \vdash \langle \sigma, env, fenv, Seq(s_0, s_1) \rangle \Rightarrow \langle \sigma', env', fenv, error \rangle} \quad \text{(EVAL-STT-SEQ2)}$$

$$FRWprogram \stackrel{\text{def}}{=} \left(Seq\ s_0\ \left(Seq\ s_1\ \left(\ldots \left(Seq\ s_{n-1} (Seq\ s_n\ Snil)\right)\right)\right)\right), (s_i \neq Seq(*), i \in \mathbb{N}) \quad (3)$$

According to EVAL-STT-SEQ1 and EVAL-STT-SEQ2, using the standard approach to evaluate a valid *Seq* statement *s*, where $s \equiv Seq\ s_0\ s_1$, $s_0$ and $s_1$ represent two arbitrary statements, and $s_0 \neq Seq(*)$, requires that *s* be processed according to the algorithm given in Table 3, where $\mathcal{FI}$ is defined as a partial function that returns an option type to indicate success or failure.

Table 3. Standard algorithm defining $\mathcal{FI}$ evaluation

Algorithm $\mathcal{FI}$_enter_point

**Function:** *Fixpoint* $\mathcal{FI}$

**Input:** Initial optional memory state $[\![m_{state}]\!]$, current environment $env$, super environment $fenv$, initial arguments $args$, and valid $FRWprogram$;

**Output:** Final memory state signed with optional type;

$\boldsymbol{Step_0}$: if $env_{check}(env, fenv) = true$, then $set_{env}(Seq(s_0, s_1), env) \hookrightarrow Some\ env'$ and go to $\boldsymbol{Step_1}$; else, exit;

$\boldsymbol{Step_1}$: if $FRWprogram = Seq(s_0, s_1)$, then go to $\boldsymbol{Step_2}$, else, $(env.K, m_{state}, env, fenv, args, FRWprogram) \Downarrow_{P(FRWprogram)}$;

$\boldsymbol{Step_2}$: if $s_0 \neq Seq(*)$, then $\boldsymbol{let}$ $\left[\!\left[m'_{state}\right]\!\right] := \mathcal{FI}([\![m_{state}]\!], env, fenv, args, s_0)$ and go to $\boldsymbol{Step_3}$, else, throw out program;

$\boldsymbol{Step_3}$: $\mathcal{FI}\left(\left[\!\left[m'_{state}\right]\!\right], env', fenv, args, s_1\right)$

Because *FRWprograms* are guaranteed to be structural programs by applying the *Program Counter* axiom, we can directly employ a pattern matching mechanism to obtain the next instruction. However, as shown in Figure 2, *Seq* is also a constructor of the *statement* type. Therefore, $\mathcal{FI}$ must first determine whether the head statement in the current context $\Gamma_c$ is a *Seq* statement. If this is the case, $\mathcal{FI}$ must evaluate the statement $s_0$ stored in *Seq* again by recursively calling itself. In higher-order theorem proving assistants, such as Coq, a recursive function will create a new logic proof context $\Gamma'$ each time a $\beta$-reduction is applied. Therefore, the current $\mathcal{FI}$ evaluation belongs to $\Gamma_c$, and the expression $\mathcal{FI}([\![m_{state}]\!], env, fenv, args, s_0)$ in $\boldsymbol{Step_2}$ of the algorithm in Table 3 belongs to $\Gamma'$. Most specification languages, such as Gallina, are a type of non-Turing-complete *FPL* that treats functions as "first-class citizens" [23], and these specification languages therefore have no mutable state like that usually applied in imperative programming languages to store the result generated in $\Gamma'$. The standard solution for addressing this condition in functional programming is to employ a *let* expression to connect contexts $\Gamma_c$ and $\Gamma'$, and temporarily store the new memory state $\sigma'$ generated by the process of executing $s_0$. In this way, $\sigma'$ can be taken as the initial memory state in the next iteration cycle for executing $s_1$. However, implementing $\mathcal{FI}$ according to the standard approach given in Table 3 includes a hidden problem, in that the actual order of expression evaluation is the opposite to that expected. Specifically, if *FRWprogram* is a statement stream connected by a *Seq* constructor, as defined by the expression

$$\frac{\dfrac{\begin{array}{c} M \vdash m_{state} \quad \mathcal{E}, M, \mathcal{F} \vdash FRWprogram \quad \mathcal{E}, M, \mathcal{F} \vdash pars \quad \mathcal{F} \vdash K \\ FRWprogram \xrightarrow{matches} Seq(s_{current}, s_1) \xrightarrow{matches} s_{current} \wedge s_1 \end{array}}{(env.K, [\![m_{state}]\!], env, fenv, pars, FRWprogram) \Downarrow_{P(s_{current})} \xrightarrow{yields} \left[\!\left[m'_{state}\right]\!\right]}}{\mathcal{FI}\left(\left[\!\left[m'_{state}\right]\!\right], set_{env}(env), fenv, pars, s_1\right)}, (4)$$

the ideal order of *FRWprogram* evaluation should be identical with the order in the real world, as follows:

1) get the current execution statement $s_{current}$;
2) evaluate $s_{current}$ and generate the new memory state $\left[\!\left[m'_{state}\right]\!\right]$ using $\mathcal{FI}$;
3) call the next statement $s_1$ in $\mathcal{FI}$ recursively.

As such, the ideal evaluation order is that the next instruction will not be executed until the current instruction is simplified as a normal form of the memory state.

Nonetheless, the standard solution of employing a *let* expression to obtain an ideal evaluation order encounters difficulties with respect to the evaluation strategy of lambda calculus adopted as the fundamental theory by most higher-order logic theorem proving systems. Here, the *let* expression is defined in lambda calculus as a lambda abstraction [24]. For example, $\boldsymbol{let}\, f\ x = y\ \boldsymbol{in}\ z$ means that a function $f$ is defined by $f\ x = y$, which is equivalent with the lambda expression $(\lambda\, f.z)(\lambda\, x.y)$, where $\lambda$ represents the abstraction. The formal definition of the *let* expression is defined according to the following rule.

$$\boldsymbol{let}\, x{:}\, T\ \boldsymbol{in}\, y \equiv (\lambda\, x.y) \quad \text{(LET-ABS)}$$

In addition, the *let* expression allows application and substitution to be applied to other expressions according to the respective following rules.

$$x \notin FV(y) \Longrightarrow \big((\boldsymbol{let}\, (x{:}\, T){:}= M\ \boldsymbol{in}\, y) \Leftrightarrow (\boldsymbol{let}\, x{:}= M\ \boldsymbol{in}\, y) \equiv (\lambda\, x.y)\ M\big) \quad \text{(LET-APP)}$$

$$x \notin FV(y) \Longrightarrow (\boldsymbol{let}\, (x{:}\, T){:}= M\ \boldsymbol{in}\, y) \equiv (\lambda\, x.y)\ M \Longrightarrow y[x \coloneqq M] \quad \text{(LET-SUB)}$$

In rule LET-APP, if $x \notin FV(y)$, where $\forall\, E.FV(E)$ represents the free variable set of expression $E$, then expression $M$ can be applied to expression $(x{:}\, T)$ bound in expression $y$. According to the substitution rule of lambda calculus [25], we can simplify rule LET-APP to obtain rule LET-SUB. Thus, the computational formal semantics given in Table 3 can be abstracted as Table 4, and, according to rules LET-ABS, LET-APP, and LET-SUB, the implementations in Tables 3 and 4 are identical.

Table 4. Abstract definition of $\mathcal{FI}$ evaluation in lambda form

$$
\begin{aligned}
&\boldsymbol{fix}\, \mathcal{FI} \equiv \\
&\qquad \lambda\, (\mathcal{FI}{:}\ option\ memory \to list\ value \to Env \to Env \to statement \to option\ memory). \\
&\quad \lambda\, (s{:}\ statement).\lambda\, (input{:}\ list\ value).\lambda\, (env, fenv{:}\ Env).\lambda\, (m_{state}{:}\ memory).. \\
&\quad \{|\ true \longmapsto \\
&\qquad \{|\ Some\ env' \longmapsto \\
&\qquad\quad \{|\ Seq\ s_0\ s_1 \longmapsto \\
&\qquad\qquad \Big(\lambda\, \big(m'_{state}{:}\ memory\big).\mathcal{FI}\big(m'_{state}, env', fenv, input, s_1\big)\Big) \\
&\qquad\qquad\qquad (\mathcal{FI}(m_{state}, env, fenv, input, s_0)); \\
&\qquad\quad _ \longmapsto Some\ m_{state}\ |\}.s; \\
&\qquad None \longmapsto Some\ m_{state}\ |\}.set_{env}(env) \\
&\quad false \longmapsto Some\ m_{state}\ |\}.\big(env_{check}(env, fenv)\big)
\end{aligned}
$$

As discussed, $\mathcal{FI}$ is essentially a large recursive function written in a specification language. Therefore, according to rules LET-APP and LET-SUB, the *let* expression for sub-statement $s_0$ given in formula (5) below can be converted into formula (6), and then evaluated as formula (7) using the evaluation tactic *simpl* or *cbn* provided by the Coq *tactic* mechanism.

$$\boldsymbol{let}\, m'_{state}{::}= \mathcal{FI}\ (m_{state}, env, fenv, parss, s_0)\ \boldsymbol{in}\, \mathcal{FI}\ \big(m'_{state'}, env, fenv, parss, s_1\big) \tag{5}$$

$$\Big(\lambda\, \big(m'_{state}{:}\ memory\big).\mathcal{FI}\ \big(m'_{state}, set_{env}(env), fenv, input, s_1\big)\Big)\ (\mathcal{FI}\ (m_{state}, env, fenv, input, s_0)) \tag{6}$$

$$\Big(\mathcal{FI}\ \big((\mathcal{FI}\ (gas, m_{state}, input, set_{env}(env), fenv, s_0)), env, fenv, input, s_1\big)\Big) \tag{7}$$

Here, we can determine that the expression $\big(\mathcal{FI}(m_{state}, env, fenv, input, s_0)\big)$ is applied directly instead of unfolding $\mathcal{FI}$ and reducing this expression to a normal form as a new memory state. Therefore, the ideal evaluation order is violated. As such, the root cause of results like formula (7) is the evaluation order of lambda calculus. In Coq, the evaluation tactics *cbn* and *simpl* adopt the *call-by-name* evaluation strategy [26], which means that the $\lambda$-expressions under lambda abstraction will not be reduced, and the arguments to a function call will not be evaluated, even though the $\lambda$-expressions are not normal forms. A simple example of the *call-by-name* evaluation strategy is illustrated in Table 5, which demonstrates that the expression $\big(\lambda\, y{:}\ int.(\lambda\, x{:}\ int.y + x)\big)\ (1+3)$ cannot be reduced to $\lambda\, x{:}\ int.4 + x$ using this strategy. The only method of simplifying the expression $\lambda\, x{:}\ int.1 + 3 + x$ is to specify $x$ in such a way as to make the expression free of $\lambda$-expressions. Moreover, according to [27], the evaluation tactics *cbn* and *simpl* will attempt first to apply $\beta$-reduction and $\iota$-reduction, and will then attempt to apply $\sigma$-reductions if necessary. Notice that only transparent constants whose identifier can be reused in the recursive calls are possibly unfolded by *simpl* and *cbn*. Accordingly, the evaluation process is illustrated in Table 6. We note from the table that the logic expression generated by the recursive function will not be simplified until the entire $FRWprogram$ connected by the *Seq* constructor has been unfolded completely, which yields the following expression:

$$\left(\mathcal{FI}_n\left(\mathcal{FI}_{n-1}\left(\dots\left(\mathcal{FI}_1(\mathcal{FI}_0(m_0, s_0,*))\right)\right)\right)\right), \quad (8)$$

where $\mathcal{FI}_i$ $(i \in \mathbb{N})$ represents the $i^{\text{th}}$ recursive call of $\mathcal{FI}$, and the wildcard * represents irrelevant parameters. Here, all *let* expressions of $\mathcal{FI}(m_{state}, FRWprograms,*)$ have been applied and $\lambda$-abstraction has been eliminated. Therefore, expression (8) is free of $\lambda$-expressions, and can be unfolded and simplified from the outside to the inside.

Table 5. Simple example of a *call-by-name* evaluation strategy

| |
|---|
| $\big(\lambda\, y: int. (\lambda\, x: int. y + x)\big)\ (1 + 3)$ |
| $\Rightarrow_\beta\ \big(\lambda\, y: int. (\lambda\, x: int. y + x)\big)\ [y \coloneqq (1 + 3)]$ |
| $\Rightarrow_\beta\ \lambda\, x: int. 1 + 3 + x$ |
| $\not\Rightarrow_\beta\ \lambda\, x: int. 4 + x$ |

Table 6. Iterations associated with $\mathcal{FI}$ evaluation

| |
|---|
| $\mathcal{FI}(m_{state}, FRWprograms,*)$ |
| $\xRightarrow{unfold} \mathcal{FI}\left(m_{state}, \left(Seq\ s_0\ \left(Seq\ s_1\ \left(\dots\left(Seq\ s_{n-1}(Seq\ s_n\ Snil)\right)\right)\right)\right),*\right)$ |
| $\overset{cbn}{\Rightarrow}_{\beta\iota\sigma} \mathcal{FI}\left(\mathcal{FI}(m_{state}, s_0,*), \left(Seq\ s_1\ \left(\dots\left(Seq\ s_{n-1}(Seq\ s_n\ Snil)\right)\right)\right),*\right)$ |
| $\overset{cbn^*}{\Rightarrow}_{\beta\iota\sigma} \cdots$ |
| $\overset{cbn}{\Rightarrow}_{\beta\iota\sigma} \mathcal{FI}\left(\mathcal{FI}\left(\dots\left(\mathcal{FI}(\mathcal{FI}(m_{state}, s_0,*), s_1,*)\right)\right), \left(Seq\ s_{n-1}(Seq\ s_n\ Snil)\right),*\right)$ |
| $\overset{cbn}{\Rightarrow}_{\beta\iota\sigma} \mathcal{FI}\left(\mathcal{FI}\left(\left(\dots\left(\mathcal{FI}(\mathcal{FI}(m_{state}, s_0,*), s_1,*)\right)\right), s_{n-1},*\right), (Seq\ s_n\ Snil),*\right)$ |
| $\overset{cbn}{\Rightarrow}_{\beta\iota\sigma} \mathcal{FI}\left(\mathcal{FI}\left(\left(\dots\left(\mathcal{FI}(\mathcal{FI}(m_{state}, s_0,*), s_1,*)\right)\right), s_{n-1},*\right), s_n,*\right)$ |

According to the above analysis, the result cannot be simplified and evaluated in the actual execution process until all recursive calls of $\mathcal{FI}$ are completed. Therefore, the actual evaluation order is given as follows:

$$\frac{\dfrac{M \vdash m_{state} \quad \mathcal{E},M,\mathcal{F} \vdash FRWprogram \quad \mathcal{E},M,\mathcal{F} \vdash par \quad \mathcal{F} \vdash K}{FRWprogram \xrightarrow{matches} Seq(s_{current}, s_1) \xrightarrow{matches} s_{current} \wedge s_1} \quad set_{args}(par) \hookrightarrow par'}{\mathcal{FI}\left(\mathcal{FI}(\llbracket m_{state} \rrbracket, env, fenv, par, s_{current}), env, fenv, par', s_1\right)}. \quad (9)$$

This order can be explicitly stated as follows:

1) get the current execution statement $s_{current}$;
2) call the next statement $s_1$ in $\mathcal{FI}$ recursively with the function call of $s_{current}$;
3) evaluate the entire *FRWprogram* and generate the final memory state $\llbracket m_{final} \rrbracket$ with $\mathcal{FI}$.

Obviously, in the actual evaluation process, $\mathcal{FI}$ takes the entire *FRWprogram* rather than a single statement as an evaluation unit. Thus, the information generated from the iterations cannot be directly simplified in normal form, and the volume of information of the current context will be too great to simplify. This condition can be abstracted as follows.

$$infor_{size} \equiv size(FRWprogram) \quad (10)$$

In addition, the above analysis indicates that an arbitrary statement within an *FRWprogram* composed of *n* statements can be evaluated after 2*n* iterations. Thus, the average number of iterations is $n * \frac{2n}{n}$. Hence, the time complexity of this process is $O(n)$.

The results presented in Figure 5 verify that the execution process in Coq is identical with the above analysis. Accordingly, the space resource of higher-order theorem proving assistants, such as Coq, will be consumed by a large volume of non-normal form logic expressions like that given in (8). Therefore, the size of an *FRWprogram* directly decreases the evaluation efficiency of the proof engine, and may even result in the failure of symbolic execution due to overload.

```
1 subgoal
pump, pump_val : nat
m, m', m0, m1, m2, m3 : memory
z, blc, gs : Z
l : list address
o : option address
a0 : address
d : dnum
s : statement
H5 : S (S (S (S (S pump)))) > 100
H6 : pump_val > 100
______________________________________(1/1)
test pump pump_val
  (modifiy_0x00000021
     (modifiy_0x00000020
        (modifiy_0x0000001f
           (modifiy_0x0000001e
              (modifiy_0x0000001d
                 {|
                 m_init := initData;
                 m_send := initData;
                 m_send_re := initData;
                 m_msg := Str_type _0xmsg
                            (str_mem Taddress (Nvar
```

Figure 5. Embedded execution result of FEther in Coq

To further clarify this issue, we define an abstract recursive type *ident* in Figure 6. Here, the identifier *ident* is its name and *sort* is its type. The identifiers $base_0$ to $base_n$ are the names of the *ident* recursive base constructors, and $cons_i$ is the rule that reduces all other cases toward the base constructors. The binders $binder_0$ to $binder_n$ are the quantifiers (such as $\forall$ and $\exists$), and $\bar{\tau}$ represents the type set of other inductive types. These terms are optional, which is indicated by placing the terms within square brackets. An inspection of Figure 6 reveals that $Seq: statement \to statement \to statement$ is obviously a special case of *ident*, in that $cons_i$: $[[binder_i]\ \bar{\tau_i}] \to ident \to ident \to [\ldots] \to ident$ is specified as the form $cons_{seq}: ident_{statement} \to ident_{statement} \to ident_{statement}$. As discussed above, the absence of a mutable state in most *FPLs* requires that, if the parameters in the constructor, which must be evaluated in $\Gamma_c$, and are therefore denoted as base parameters, cannot be evaluated, we can only transmit the base parameters into the next recursive circle or discard them. However, it is also clear that the transmission of base parameters is limited because, if the current recursion period transmits $n$ base parameters into the next recursion period, then the next recursion period must in turn transmit $2n$ base parameters. Therefore, the $m^{th}$ recursion period will need to transmit $m * n$ base parameters into the next recursion period. The strict type system employed by higher-order theorem proving assistants requires that the parameters and the respective types of each function must be defined explicitly. Hence, it is impossible to transmit the remaining base parameters into the next recursion period.

$$
\begin{aligned}
&Inductive\ ident : [[binder]\ \bar{\tau}] \to\ sort :=\\
&\qquad |\ base_0 : [[binder_0]\ \bar{\tau}_0] \to ident\\
&\qquad |\ base_1 : [[binder_1]\ \bar{\tau}_1] \to ident\\
&\qquad \ldots\\
&\qquad |\ cons_i : [[binder_i]\ \bar{\tau_i}] \to ident \to ident \to [\ldots] \to ident\\
&\qquad \ldots\\
&.
\end{aligned}
$$

Figure 6. An abstract data type that illustrates the *call-by-name termination* (CBNT) problem

Unfortunately, most higher-order theorem proving assistants adopt *call-by-name* as their essential evaluation strategy. Thus, CBNT is a common problem in all studies where researchers have followed the standard approach for designing a computational proof engine in these higher-order theorem proving assistants to evaluate formal programs at the code level, or when researchers have defined very large recursive functions to evaluate recursive datatypes like the abstract datatype given in Figure 6.

**Problem 2** (***Information redundancy explosion***) The cause of IRE is primarily the result of a common programming style, and IRE is exacerbated by CBNT. In most cases, the major component of functions written in *FPLs* consists of conditional statements defined by a pattern matching mechanism. In general, the programming style involves defining these conditions explicitly in the function body rather than encapsulating these conditions as a function interface. In fact, higher-order theorem proving assistants actually encourage users to apply this type of programming style in programs. To this end, proof assistants provide wildcard syntactic sugar to simplify the manual definition, and the built-in interpreter will automatically fill the wildcard during the evaluation process. A simple example of this process is given in Table 7 for an inductive type *T* that has three constructors *A*, *B*, and *C*. Here, wildcards have been used to define a function on the left side of the table, and the trusted core of the proof assistant automatically fills the wildcards, as shown on the right side of the table.

Table 7. Simple example of pattern matching definition in Coq. The definition with wildcards is given on the left, while the actual definition completed by the core of Coq is given on the right.

| | |
|---|---|
| $Definition\ test\ (a{:}\ T){:}\ bool \coloneqq$<br>$match\ a\ with$<br>$\mid A \Rightarrow true$<br>$\mid _ \Rightarrow false$<br>$end.$ | $Definition\ test\ (a{:}\ T){:}\ bool \coloneqq$<br>$match\ a\ with$<br>$\mid A \Rightarrow true$<br>$\mid B \Rightarrow false \mid C \Rightarrow false$<br>$end.$ |

In addition, as discussed above, the evaluation tactics *cbn* and *simpl* can only be successfully applied to transparent logic expressions free of $\lambda$-abstraction [27]. In other words, proof assistants like Coq only apply $\beta$-reduction rules for $\lambda$-expressions when the top-level structure [28] of the terms of the $\lambda$-expressions is deconstructed as specific constructors without $\lambda$-abstraction. This is illustrated by the simple example given in Table 7 based on the process. Here, researchers seeking to prove the theorem $\forall\ (a{:}\ T), test\ a\ \wedge false = false$ in a proof assistant like Coq requires the completion of two basic steps. First, according to higher-order lambda calculus, the theorem $\forall\ (a{:}\ T), test\ a\ \wedge false = false$ is equivalent with $\lambda\ (a{:}\ T), test\ a\ \wedge false = false$. Hence, the $\lambda$-abstraction should be specified with a term of type $T$. To avoid confusion, the specific term is bound with the name $a_0$. Therefore, $\lambda\ (a{:}\ T), test\ a\ \wedge false = false$ is transformed as $(a_0{:}\ T) \vdash test\ a_0\ \wedge false = false$. Second, the $test$ will be unfolded in $\Gamma_c$ as follows.

$$(match\ a_0\ with \mid A \Rightarrow true \mid _ \Rightarrow false\ end) \wedge false = false \quad (11)$$

Because $a_0$ is a top-level structure term, the expression given in (11) cannot be simplified directly. Therefore, $a_0$ should be deconstructed according to its constructors *A*, *B*, and *C*, which generates three sub-goals of the proof, respectively. Each sub-goal can be proven easily, where the sub-goal of *A* can be evaluated as true, while the sub-goals of B and C can be evaluated as false.

The evaluation and verification process illustrated by the above example is equivalent with that conducted by the proposed FSPVM-E. As discussed previously, FEther is a recursive function that is constructed entirely based on the GERM framework in Coq, so the sum total of logic information, such as $\lambda$-expressions and proof terms, must be evaluated and verified in $\Gamma_c$ by the trusted core of the proof assistant, which contrasts with the process conducted in a real world virtual machine using actual hardware. Therefore, during the ideal process of evaluating *FRWprograms* and generating the new memory state $[\![ m'_{state} ]\!]$ using $\mathcal{FI}$, the entire $\mathcal{FI}$ structure is treated as a function definition, and unfolded in $\Gamma_c$.

However, this process includes a hidden problem. A higher-order theorem proving system must display all logic information in the same level context to maintain logical consistency. In other words, all logic expressions defined explicitly according to the standard programming style, rather than being encapsulated as a function interface, will be displayed in $\Gamma_c$ by the proof assistant. Therefore, as discussed above, all wildcards in the definition of $\mathcal{FI}$ will be automatically filled, and pattern matching conducted without definition encapsulation will be unfolded in $\Gamma_c$. However, because the logic information in $\Gamma_c$ cannot be simplified and evaluated by tactics prior to deconstructing the top-level structure terms of the $\lambda$-expressions as specific constructors without $\lambda$-abstraction, very large formal programs like the non-optimized version of FEther will generate more than 5000 lines of logic expressions in $\Gamma_c$ during a single recursive cycle.

In addition, as shown in Table 3, the first step is to recursively apply $\mathcal{FI}$ for the next statement and nestedly unfold $\mathcal{FI}_i$ in $\Gamma_c$, rather than simplifying the terms of $\Gamma_c$. Therefore, pattern matching in the function body cannot be simplified promptly, and all definitions in all branches will be unfolded in $\Gamma_c$. Unfortunately, most branches need not be unfolded in $\Gamma_c$ because the natural deduction systems of higher-order theorem proving theory can prune irrelevant branches. Specifically, according to the inductive datatype principle, all constructors of a datatype are mutually exclusive with each other. This is illustrated by expression (11), where it is impossible to generate constructors *A* and *B* into the matching guard of *test* simultaneously. Hence, a single branch at most of a pattern matching process can be derived by a deterministic state $\Phi_i$ of $\Gamma_c$ during the forward reasoning deduction process, and other branches will be filtered out. This is abstracted as follows.

$$\frac{\Gamma_c(\Phi_i) \rhd (B_0, B_1, \dots, B_n)}{[\![B_i]\!]} \quad (12)$$

As such, the definitions belonging to these irrelevant branches will not be unfolded in $\Gamma_c$. This condition is illustrated by the test shown in Figure 3. Here, when $if_{throw}$ is evaluated in Coq, the statement *s* is bound with a universal quantifier, which is an unknown top-level value term. Therefore, for *s*, which has type *statement*, all its possible pattern matching combinations, including all branches and sub-branches, will be explicitly presented in the current logic context of Coq. In Figure 7, this process generated 10,736 lines of logic information in the current context. However, according to the logic process of $if_{throw}$, if the condition expression is true, the result of $if_{throw}$ is fixed as $init_m$ and independent of *s*. As such, the frustrating truth is that most of the logic information details in the 10,736 lines are redundant and irrelevant.

Therefore, although the conditional statement given in Example 1 is very simple, *s* is a top-level term that cannot be simplified directly, and, because *s* is unfolded first, but is evaluated last, CBNT exacerbates the problem of IRE, resulting in an excessive volume of unnecessary logic expressions that must be handled by the trusted core of the proof assistant.

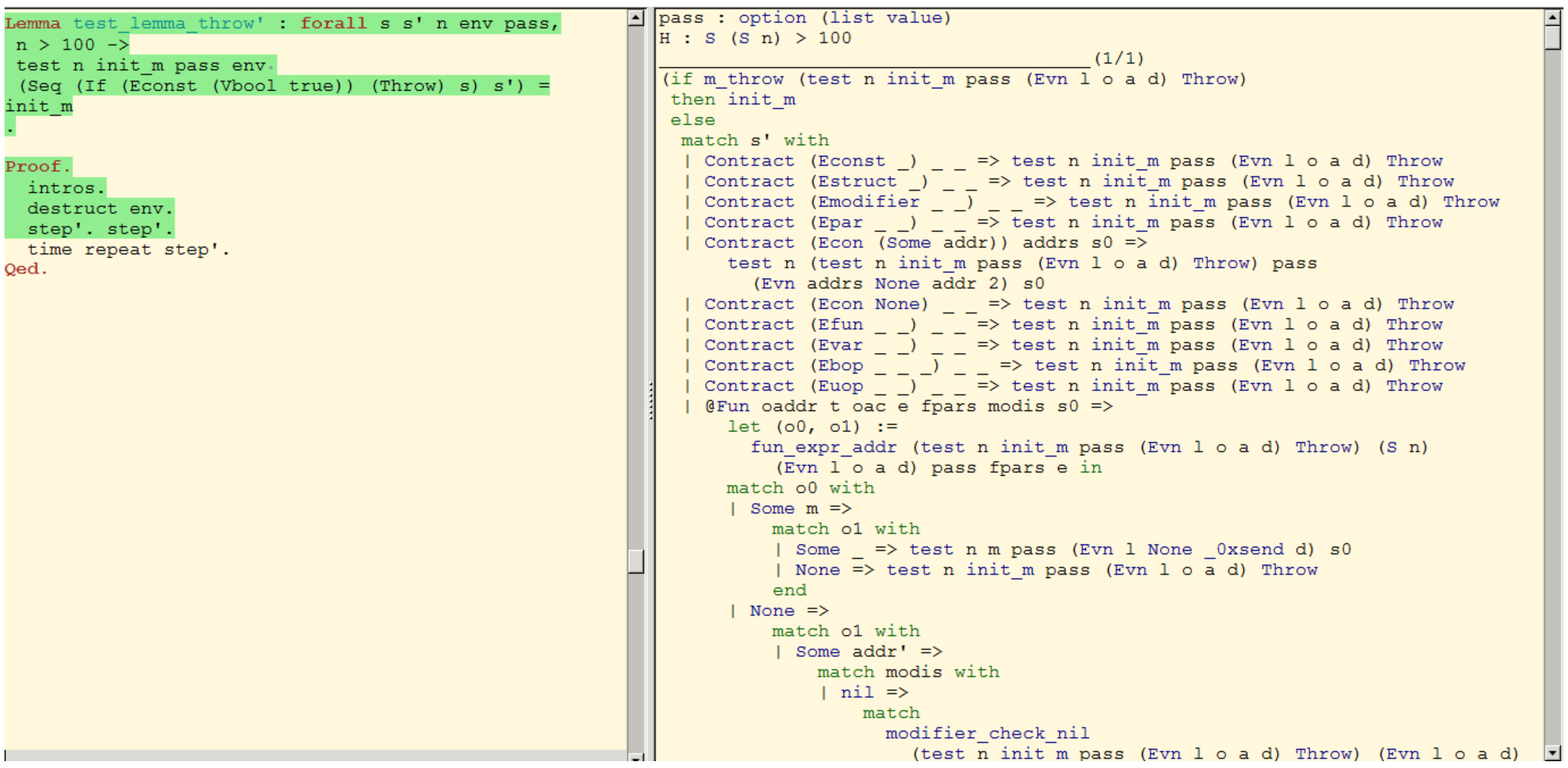

Figure 7. Example illustrating irrelevant logic information unfolding in the proof context

We also note that, in addition to the size of *FRWprograms*, the size and complexity of the $\mathcal{FI}$ structure are also elements influencing the computational efficiency because the function evaluation process typically generates a large volume of logic information combinations, particularly for a very large recursive function like $\mathcal{FI}$. Therefore, we redefine formula (10) as follows.

$$infor_{size} \equiv \big(num(FRWprogram) * size(\mathcal{FI})\big) + size(FRWprogram) \quad (13)$$

As mentioned previously, the number of recursion steps required by $\mathcal{FI}$ is equivalent to the number of statements in an *FRWprogram*, so $num(FRWprogram)$ represents the nested depth of iterations. In addition, $size(\mathcal{FI})$ is mainly determined by the branches of the matching definition. Thus, it can be summarized as follows.

$$size(\mathcal{FI}) \equiv \left(c_{nosub_1} + c_{sub_1} * \left(c_{nosub_2} + c_{sub_2} * \left(\dots\left(c_{nosub_i} + c_{sub_i} * (\dots)\right)\right)\right)\right) \quad (14)$$

Here, $c_{nosub}$ represents the average number of constructors without sub-branches, $c_{sub}$ represents the average number of constructors with sub-branches, and $c_{sub_j}$ and $c_{nosub_j}$ represent the number of sub-branches of $c_{sub_i}$ $(i, j \in \mathbb{N}, 0 \le i \le j)$. Obviously, formula (13) can be summarized as follows:

$$infor_{size} \equiv \left(num(FRWprogram) * \left(\textstyle\sum_{i=1}^{n} c_{nosub_i} * c_{sub_{i-1}}!\right)\right) + size(FRWprogram), \quad (15)$$

where we use $c_{sub_n}!$ to represent the factorial expression $c_{sub_1} * c_{sub_2} * \dots * c_{sub_n}$, and $c_{sub_0}! = 1$.

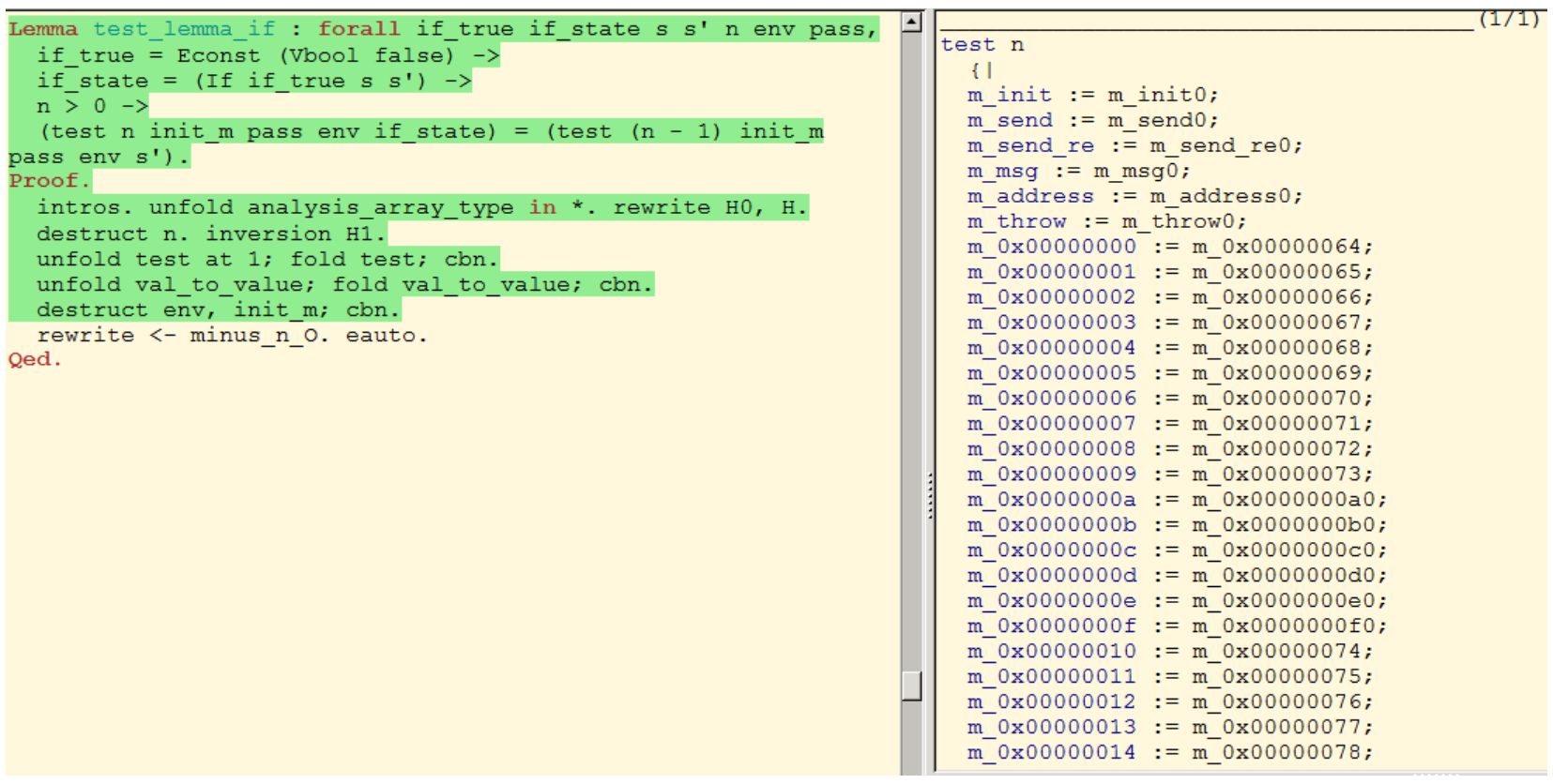

Figure 8. Example of the unfolding of memory space terms in the proof context

In addition, for the FSPVM-E framework proposed herein, FEther is based on our GERM framework, which is defined as a large data structure with a memory space abstracted as a special record type *memory*. Therefore, a memory state $\sigma$ is treated as a very large record value with a $memory$ type in the trusted core of Coq. As such, $\sigma$ is treated in the Coq computation process as an unknown top-level structure that must be unfolded and deconstructed according to its specific constructors during its evaluation process, and all information stored in the current memory state will be shown in the current context while waiting to be simplified. This is illustrated by the Coq evaluation process shown in Figure 8 for a simple example. According to the calculus of inductive construction, all identical value terms in the same level context must be deconstructed or unfolded simultaneously to maintain consistency. This is defined for FEther as follows.

$$size(\mathcal{F}Ether) \equiv c_{nosub_1} + (m_1 + a_1) * m_{size} + c_{sub_1} * \left(c_{nosub_2} + (m_2 + a_2) * m_{size} + c_{sub_2} * \left(\ldots\left(c_{nosub_i} + (m_i + a_i) * m_{size} + c_{sub_i} * (\ldots)\right)\right)\right) \quad (16)$$

Here, the values such as memory space terms *m*, which are a kind of special large value, and memory address terms *a* in all branches will be unfolded simultaneously. We use $m_i$ and $a_i$ to represent the average number of memory space value terms and address value terms in a branch, respectively. Because types *memory* and *address* have the same number of constructors, their sizes are identically defined as $m_{size}$. Therefore, the size of FEther in the current logic context during the evaluation process is summarized as follows.

$$infor_{FEther} \equiv \left(num(FRWprogram) * \left(\sum_{i=0}^{n}\left(c_{nosub_i} + (m_i + a_i) * m_{size}\right) * c_{sub_{i-1}}!\right)\right) + size(FRWprogram) \quad (17)$$

Clearly, this represents an exponential growth in the volume of information in a given logic context, which results in a very large volume of logic information that must be evaluated in a current proving context by proof assistants. Therefore, the burden of computation is very large, even if *FRWprogram* is a very simple code segment.

Finally, because IRE is caused by the programming style and the basic features of higher-order proof assistants, we note that normal large programs developed in higher-order theorem proving assistants will also cause this problem. Hence, the final $infor_{size}$ can be abstracted as follows.

$$infor_{size} \equiv \left(num(FRWprogram) * \left(\sum_{i=0}^{n}\left(es_i + ds_i + c_{nosub_i} + [r_0 \quad \cdots \quad r_i] * \begin{bmatrix} size_0 \\ \ldots \\ size_i \end{bmatrix}\right) * c_{sub_{i-1}}!\right)\right) + size(FRWprogram) \quad (18)$$

Here, $r_0$ to $r_i$ represent the number of values constructed by different datatypes, and $size_0$ to $size_i$ represent the number of constructors for each respective datatype. In addition, $infor_{size}$ also contains basic expressions and definitions that can be evaluated directly, and the average number of these basic expressions and definitions are defined as $es_i$ and $ds_i$, respectively.

Problem 3 (***Concurrent reduction***) The present implementation of FSPVM-E seeks to combine the advantages of model checking and theorem proving in proof loops and avoid halting problems. To this end, we have employed Bounded Model Checking (BMC) [29] in EVI by allowing FEther to unfold and execute $FRWprograms$ only *K* times. This approach, which is denoted as *fuel* or *pump*, avoids functions invoking infinite loops, and corresponds to the *gas* approach employed by Ethereum [31], where the evaluation process of semantics are halted if the level of *gas* fails to pass the gas checking function.

In the standard FEther design, an equivalent *K*-limitation is employed to limit the symbolic execution in the statement, expression, and value semantic layers simultaneously, rather than using different *K* values in different layers. However, as discussed previously, all identical value terms in the same context must be deconstructed or unfolded simultaneously to maintain consistency. Therefore, the value of *K* will be modified and shared among all layers, and the layers that await the execution result will also be forced to be unfolded, as illustrated in Figure 9. While this process will not cause data corruption owing to the forward reasoning of higher-order theorem proving, it will cause IRE to occur more frequently because $c_{nosub_i}$ will become $c_{sub_i}$ in formula (14).

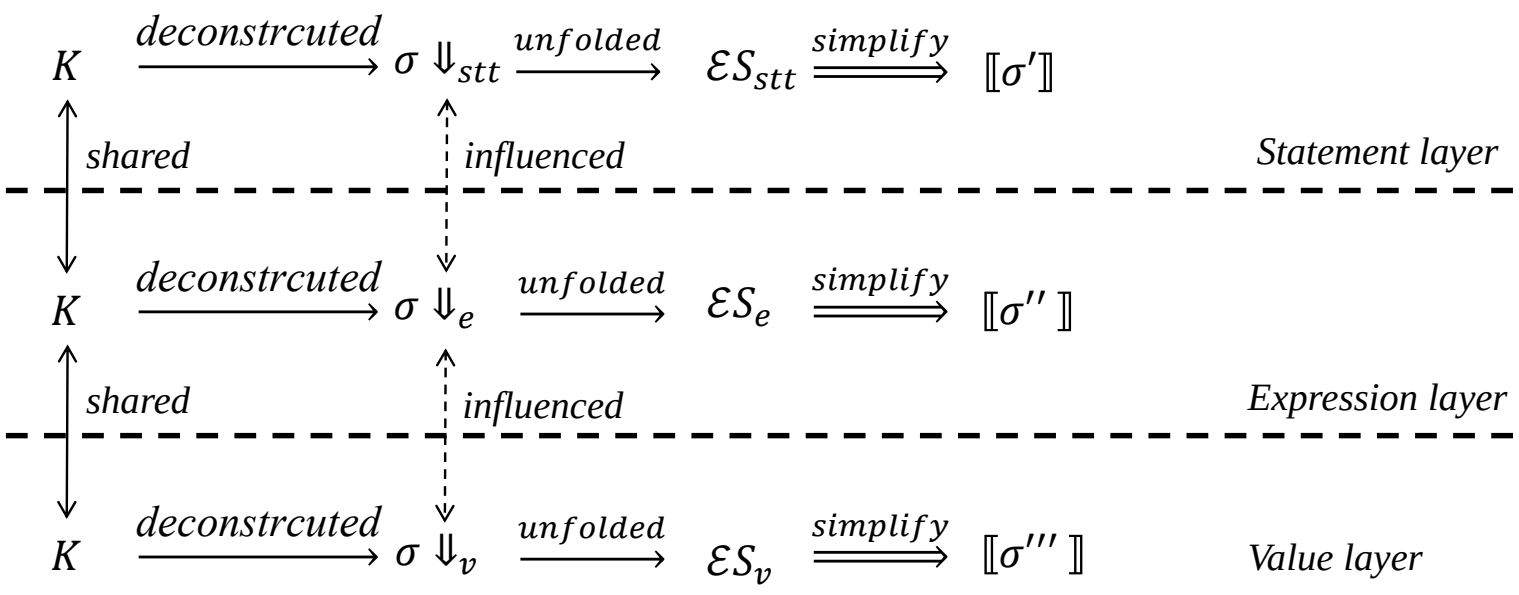

Figure 9. Illustration of how the shared *K*-limitation influences all layers simultaneously

The simple example given as Example 1 above is employed to illustrate the CR problem in Figure 10. Here, because the different layers share an equivalent *K*-limitation (i.e., *pump*), when *pump* in the first matching guard, which belongs to the statement layer, is deconstructed, the function *fun_expr_addr*, which takes *pump* as its limitation in the *Fun* branch, will be unfolded at the same time. However, the second matching guard cannot simplify the *stt* in this step. Therefore, *fun_expr_addr* also cannot be simplified after unfolding, and the logic information will remain in the proof context. Unfortunately, this condition will exist in other branches, but most of the branches are irrelevant after the *stt* is deconstructed. This means that a large volume of irrelevant logic information will be generated before the *stt* is deconstructed.

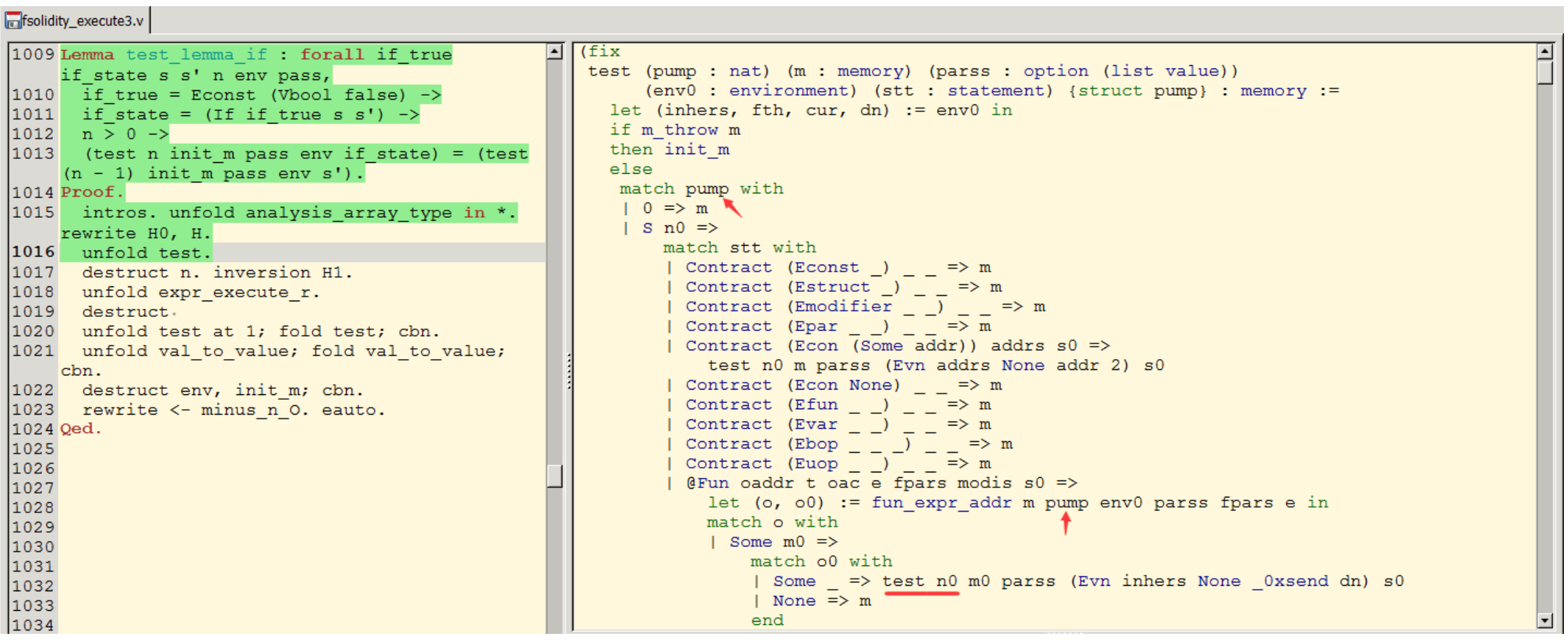

Figure 10. Example based on Example 1 illustrating the *concurrent reduction* (CR) problem

Finally, it should be noted that all functions developed according to BMC in higher-order theorem proving assistants will also suffer from CR.

## 5. Optimization

### 5.1 Solutions for each problem

According to the above analysis, $\mathcal{FI}$, like all large programs written by a specification language in higher-order theorem proving assistants that adopt the *call-by-name* strategy and BMC, will be subject to CBNT, IRE, and CR problems. We therefore present three general optimization methods for addressing each of these problems respectively.

***Solution*** 1 (***call-by-name termination***) To clarify the present discussion, we summarize the cause of CBNT as follows. If the *Seq* cell constructor is defined explicitly in $\mathcal{FI}$, the execution of $\mathcal{FI}$ requires that the pattern matching mechanism successively obtain the *Seq* result and $s_0$ in the current context $\Gamma_c$. Subsequently, CBNT occurs during the evaluation process where $s_0$ is evaluated in $\Gamma'$ and the result is bound in $\Gamma_c$ by the *let* expression. However, this summary indicates that the CBNT problem can be solved directly if $s_0$ is directly evaluated in $\Gamma_c$ rather than in $\Gamma'$. One of the available solutions is that *Seq* can be defined implicitly. Specifically, this solution involves removing the *Seq* constructor from the *statement* inductive datatype, and defining the sequence statement implicitly using the *list* datatype. The new semantics of the proposed sequence statement are defined in rules NEW-STT-SEQ1 and NEW-STT-SEQ2 below.

$$\frac{M \vdash \sigma \qquad \mathcal{E},M,\mathcal{F} \vdash s\colon list\ statement' \qquad \sigma \Downarrow_{s.head}\left(\sigma',normal\right) \qquad env_{check}(env,fenv) \hookrightarrow true \wedge set'_{gas}(s.head::nil,env) \hookrightarrow Some\ env'}{\mathcal{E},M,\mathcal{F} \vdash \langle \sigma,env,fenv,s\rangle \Rightarrow \langle \sigma' \Downarrow_{s.tail},env',fenv,normal\rangle} \quad \text{(NEW-STT-SEQ1)}$$

$$\frac{M \vdash \sigma \qquad \mathcal{E},M,\mathcal{F} \vdash s\colon list\ statement' \qquad \sigma \Downarrow_{s.head}\left(\sigma',error\right) \qquad env_{check}(env,fenv) \hookrightarrow true \wedge set'_{gas}(s.head::nil,env) \hookrightarrow Some\ env'}{\mathcal{E},M,\mathcal{F} \vdash \langle \sigma,env,fenv,s\rangle \Rightarrow \langle \sigma',env',fenv,error\rangle} \quad \text{(NEW-STT-SEQ2)}$$

Here, we use *list* to connect the value constructed by the new statement type *statement'*, which does not have the *Seq* constructor. Because no other semantics are modified, the $\sigma \Downarrow_{stt}$ process is still adopted in the new sequence statement semantics. Based on the above definitions, the equivalence between the standard sequence statement semantics and the new semantics is proven by the Theorem *Sequence Equivalence* given below. However, we first require an intermediate function $\mathcal{T}$ to equivalently translate an *FRWprogram* from the *statement* type to the *statement'* type. The abstract definition of the translation function is assigned as $\mathcal{T} :: statement \rightarrow list\ statement'$. Its expected function is redefining, where, if *FRWprogram* is $\left(Seq\ s_0\ \left(Seq\ s_1\ \left(\dots\left(Seq\ s_{n-1}\left(Seq\ s_n\ snil\right)\right)\right)\right)\right)$, the new *FRWprogram* should be $s_0 :: s_1 :: \cdots :: s_{n-1} :: s_n :: nil$. The correctness of $\mathcal{T}$ can be guaranteed by the Lemma *Translation* given below, where we assume that the implementation of $\mathcal{T}$ is correct.

***Lemma*** (*Translation*) $\forall\ (K\colon nat)\ (FRWprogram\colon statement)\ (m\colon memory)\ (env, fenv\colon environment)$.

***Judgment*** 1: $\left(\forall\ (s_0\ s_1\colon statement), FRWprogram = Seq(s_0, s_1) \rightarrow s_0 \neq Seq(*)\right) \rightarrow$.

***Goal***: $FRWprogram = \mathcal{T}^{-1}(\mathcal{T}(FRWprogram))$.

***Theorem*** (*Sequence Equivalence*) $\forall$ $(K: nat)$ $(FRWprogram: statement)$ $(m: memory)$ $(env, fenv: environment)$.

***Judgment*** 1: $(\forall\ (s_0\ s_1: statement), FRWprogram = Seq(s_0, s_1) \rightarrow s_0 \neq Seq(*))$.

***Goal***: $\mathcal{E}_{stt}(m, env, fenv, FRWprogram) = \mathcal{E}_{stt}^{'}(m, env, fenv, \mathcal{T}(FRWprogram))$.

***Proof***.

An $FRWprogram$ is defined with an inductive type. Therefore, it can be inducted as the basic *Seq* statement $s_b$, where $s_b \neq Seq(*)$ and $Seq(s_0, s_1)$. According to the definition of $\mathcal{T}$, $\mathcal{T}(s_b)$ can be evaluated as $s_b :: nil$, and $\mathcal{T}(Seq(s_0, s_1))$ can be evaluated as $s_0 :: s_1^{'}$, where $s_1^{'}$ has type $list\ statement'$.

First, the proof goal above is converted for $s$ to prove $\mathcal{E}_{stt}(m, env, fenv, s_b) = \mathcal{E}_{stt}^{'}(m, env, fenv, s_b :: nil)$. For the left side, $\mathcal{E}_{stt}(m, env, fenv, s_b) = m \Downarrow_{s_b}$ according to the rule EVAL-STT-SEQ1. For the right side, $\mathcal{E}_{stt}^{'}(m, env, fenv, \mathcal{T}(FRWprogram)) = m \Downarrow_{(s_b :: nil).head} = m \Downarrow_{s_b}$ according to rule NEW-STT-SEQ1. Therefore, $\mathcal{E}_{stt}(m, env, fenv, s_b) = \mathcal{E}_{stt}^{'}(m, env, fenv, s_b :: nil)$ is true, which yields the following judgment.

***Judgment*** 2: ***Judgment*** 1 $\vdash$ $\forall\ (s: statement), \mathcal{E}_{stt}(m, env, fenv, s) = \mathcal{E}_{stt}^{'}(m, env, fenv, \mathcal{T}(s))$.

Second, the original proof goal is converted as follows.

***Goal'***: $\forall\ (s\ s_2: statement), \mathcal{E}_{stt}(m, env, fenv, Seq(s_2, s)) = \mathcal{E}_{stt}^{'}\left(m, env, fenv, \mathcal{T}(Seq(s_2, s))\right)$.

We simplify *Goal'* according to the definition of $\mathcal{T}$ as $\forall\ (s\ s_2: statement), \mathcal{E}_{stt}(m, env, fenv, Seq(s_2, s)) = \mathcal{E}_{stt}^{'}(m, env, fenv, s_2 :: s)$. For the left side of *Goal'*, $s_2 \neq Seq(*)$ according to Judgment 1. Therefore, the left side can be evaluated according to the rule EVAL-STT-SEQ1 as follows.

$$\mathcal{E}_{stt}(\mathcal{E}_{stt}(m, env, fenv, s_2), env', fenv, s) \quad \text{(H1)}$$

Similarly, we can follow the process defined in the rule EVAL-STT-SEQ2 to evaluate the right side of *Goal'* as follows.

$$\mathcal{E}_{stt}^{'}\left(\mathcal{E}_{stt}^{'}(m, env, fenv, s_2 :: nil), env', fenv, s\right) \quad \text{(H2)}$$

According to *Judgment* 2, $s$ can be specified as $s_2$. Hence, $\mathrm{H}_1: \mathcal{E}_{stt}(m, env, fenv, s_2) = \mathcal{E}_{stt}^{'}(m, env, fenv, s_2 :: nil)$. If the output state of $\mathrm{H}_1$ is an error, $\mathcal{E}_{stt}(\mathcal{E}_{stt}(m, env, fenv, s_2), env', fenv, s) = \mathcal{E}_{stt}^{'}\left(\mathcal{E}_{stt}^{'}(m, env, fenv, s_2 :: nil), env', fenv, s\right) = error$. Otherwise, $\mathrm{H}_1$ can be assigned as $m'$. Hence, the left side of *Goal'* is $\mathcal{E}_{stt}(m', env', fenv, s)$ and the right side is $\mathcal{E}_{stt}^{'}\left(m', env', fenv, s\right)$. Furthermore, $s'$ can be specified as $s$ according to *Judgment* 2. Therefore, $\mathcal{E}_{stt}^{'}\left(m', env', fenv, s\right) = \mathcal{E}_{stt}(m', env', fenv, s)$. Hence, we successfully prove the Theorem *Sequence Equivalence*.

Because the old semantics are equivalent with the new semantics, and the new semantics also satisfy the Axiom *Pointer Counter*, the new evaluation algorithm for $\mathcal{FI}$ can be redefined according to that given in Table 8 based on the rules NEW-STT-SEQ1 and NEW-STT-SEQ2. Because the semantics of all other statement types are not modified, the $\Downarrow_{P(FRWprogram)}$ process still represents the process of evaluation.

First and foremost, this modification solves the CBNT problem. This is illustrated in Table 8 by the fact that the evaluation unit at each step is a single statement rather than the entire *FRWprogram*. The *list* datatype is an individual polymorphic recursive type, so the *list* datatype can take the *statement* datatype as its parameter and be specified as a $list\ \{statement\}$ datatype whose list elements are values with the *statement* datatype. If an *FRWprogram* is a statement list, the head of *FRWprogram* is the next executed statement, and it can be evaluated by $\Downarrow_{P(s)}$ directly in $\Gamma_c$, rather than first employing $\Downarrow_{P(s)}$ to evaluate sequence statement semantics. Therefore, the version of $\mathcal{FI}$ employing the new evaluation algorithm will not be employed again to evaluate $s_0$, and this process will also not be bounded by λ-abstraction due to the *let* expression. Thus, as illustrated in Table 9, the actual evaluation order is the same as the expected order, which takes a statement as an evaluation unit. Therefore, the specification $num(FRWprogram)$ in formula (18) is simplified to the specification $num(statement)$, where $num(statement)$ can be viewed as the special case of $num(FRWprogram)$ for an *FRWprogram* with only a single statement. As such, $num(statement) = 1$. Hence, the new $infor_{size}$ is optimized as follows.

Table 8. New algorithm defining evaluation in the revised $\mathcal{FI}$ (i.e., $\mathcal{FI}'$ )

| |
|---|
| Algorithm $\mathcal{FI}'$ _enter_point |
| **Function:** *Fixpoint* $\mathcal{FI}'$ |
| **Input:** Initial *K* steps, optional initial memory state $om_{state}$, current environment *env*, super environment *fenv*, initial arguments $args$, and valid $FRWprogram'$ . |
| **Output:** Final memory state signed with optional type. |
| $\boldsymbol{Step_0}$: if $env_{check}(env, fenv) = true$, then go to $\boldsymbol{Step_1}$, else throw out program; |
| $\boldsymbol{Step_1}$: if $FRWprogram = s_0 :: s_1$, then $env' = set_{env}(s_0 :: s_1, env)$ and go to $\boldsymbol{Step_2}$, else $[\![m_{state}]\!]$; |
| $\boldsymbol{Step_2}$: if $(m, env, fenv, \varepsilon_0) \Downarrow_{P(s_0)} \xrightarrow{yields} [\![m'_{state}]\!]$, then go to $\boldsymbol{Step_3}$, else throw out program; |
| $\boldsymbol{Step_3}$: $\mathcal{FI}'\left([\![m'_{state}]\!], env', fenv, args, s_1\right)$. |

Table 9. Abstract evaluation process of an *FRWprogram* in the revised $\mathcal{FI}$ (i.e., $\mathcal{FI}'$ )

$$\mathcal{FI}'\ ([\![m_{state}]\!], FRWprograms, *)$$
$$\xRightarrow{unfold} \mathcal{FI}'\ ([\![m_{state}]\!], s_0 :: s_1 :: \cdots :: nil, *)$$
$$\overset{cbn}{\Rightarrow}_{\beta\iota\sigma} \mathcal{FI}'\ \left((env.K, [\![m_{state}]\!], env, fenv, \varepsilon_0) \Downarrow_{P(s_0)}, s_1 :: \cdots :: nil, *\right)$$
$$\overset{cbn}{\Rightarrow}_{\beta\iota\sigma} \mathcal{FI}'\ ([\![m_{state_0}]\!], s_1 :: \cdots :: nil, *)$$
$$\overset{cbn^*}{\Rightarrow}_{\beta\iota\sigma} \cdots$$
$$\overset{cbn}{\Rightarrow}_{\beta\iota\sigma} [\![m_{state_n}]\!]$$

$$\begin{aligned} &infor_{size} \\ &\equiv \big(size(statement) * num(\mathcal{FI})\big) + size(FRWprogram) \\ &\equiv \left(size(statement) * \left(\sum_{i=0}^{n}\left(es_i + ds_i + c_{nosub_i} + [r_0 \ \cdots \ r_i] * \begin{bmatrix} size_0 \\ \cdots \\ size_i \end{bmatrix}\right) * c_{sub_{i-1}}!\right)\right) + size(FRWprogram) \\ &\equiv \left(\sum_{i=0}^{n}\left(es_i + ds_i + c_{nosub_i} + [r_0 \ \cdots \ r_i] * \begin{bmatrix} size_0 \\ \cdots \\ size_i \end{bmatrix}\right) * c_{sub_{i-1}}!\right) + size(FRWprogram) \end{aligned} \tag{19}$$

Here, we note that the component $size(\mathcal{FI})$ corresponding to the first term on the right will not be recursively called, and the time complexity for evaluating a sequence statement is reduced as $O(1)$.

Second, the new definition not only reduces $infor_{size}$, but it also strengthens the typing judgment of sequence statements. This can be explained as follows. In contrast to the original definition, the new definitions NEW-STT-SEQ1 and NEW-STT-SEQ2 are not essential for defining the side condition $\forall\, s: statemnt, if\ s = Seq(s_0, s_1)\ then\ s_0 \neq Seq\left(s, s'\right)$ because the *Seq* constructor is removed from the *statement* type in the new *statement*, and the constructor is a typing parameter of *list* type, where, according to the *list* type [30], the connection constructor is $cons\ \{statement\} :: statement \rightarrow list\ statement \rightarrow list\ statement$. Therefore, if the first parameter is not a statement type, the list typing judgment aids the type-checking mechanism in the trusted core of proof assistants to locate errors, such that the side condition need not be defined in the new sequence statement typing judgment.

Of course, Solution 1 is also a generic solution for the CBNT problem in different situations. It should also be noted that, while defining pattern matching for each parameter explicitly may seem to be another available solution, this scheme will actually make the problem more serious. This can be illustrated by the abstract recursive datatype given in Figure 6. Here, the number of parameters is arbitrary, and, if the number of parameters is $n$, the function must define pattern matching explicitly for $n-1$ parameters in the function body. However, the pattern matching for each parameter is identical to the pattern matching results of all other parameters. Therefore, the new size of $\mathcal{FI}$ can be expressed as $size_{new}(\mathcal{FI}) \equiv size(\mathcal{FI}) * n$, which correspondingly increases $infor_{size}$, and makes this solution counterproductive.

Finally, the abstract recursive datatype given in Figure 6 can further illustrate how our proposed solution of defining a new higher level connection datatype is one of the best solutions for the CBNT problem. As mentioned in the discussion of Problem 1, the feature of functional programming provided by higher-order theorem proving assistants requires that the base parameters in the constructor $cons_i$: $[[binder_i]\ \overline{\tau_i}] \rightarrow ident \rightarrow ident \rightarrow [\dots] \rightarrow ident$ be evaluated in the current context; otherwise, the information will be lost. However, the $list\ \{ident\}$datatype solves this problem by replacing $cons_i$ in datatype $ident$. Moreover, the term $cons_i\ id_0\ id_1\ id_2 \dots ids$ obtained after replacement can be equivalently redefined as the term $cons\ \left(id_0\ cons\ \left(id_1\ cons\ (id_2 \dots nil)\right)\right) \oplus ids$,, where the symbol $\oplus$ represents combination.. As such, we need only evaluate the head of the $list\ \{ident\}$ in the current recursion period, and the remaining base parameters can be completely transmitted to the next recursion period within the list. Although this increases the number of recursion periods, it solves Problem 1 completely, and the correctness of the translation can be easily proven by defining a lemma like *Translation*.

***Solution*** **2** (***Information redundancy explosion***) While CBNT is solved by Solution 1, the $size(\mathcal{FI})$ component given in formula (19) remains very large, and this will import an excessive volume of logic information in $\Gamma_c$, resulting in IRE.

Under ideal conditions, only necessary information would be included in $\Gamma_c$, and the pattern matching results in $\Gamma_c$ could be simplified directly, rather than being held in $\Gamma_c$. This would eliminate the factorial term in (19), and thereby maintain a manageable volume of information in $\Gamma_c$. These ideal conditions can be achieved through prioritization, and the best prioritization scheme to achieve this end is dynamic programming based on the *call-by-name* evaluation strategy. This is an interesting finding, in that, although CBNT is caused by the evaluation order of the *call-by-name* strategy used by tactics in higher-order theorem proving assistants, this strategy can be used to solve the IRE problem. According to this strategy, the bodies of all definitions, including functions and values, are stored in their own contexts, and are not evaluated until they are needed in $\Gamma_c$. Therefore, this feature can be taken advantage of to hide information not required for use in $\Gamma_c$.

The specific process for achieving this end is illustrated in Figure 11, where the surface context $\Gamma$, which is usually $\Gamma_c$, consists only of basic expressions $\varepsilon$ whose results can be matched in the matching guard directly. The matching results, their sub-matching results, functions, and special values, such as memory states and addresses, are separately encapsulated into definitions. In particular, the super-matching results must be separated from the sub-matching results. For example, if $definition_i$ is a matching result, its sub-matching result should be separately defined in $d_l$. In this way, an optimal evaluation process for $\mathcal{FI}$ can be obtained based on dynamic programming by breaking the process into sub-processes, and then recursively evaluating the simple results of all sub-processes. This can be abstracted as follows.

$$\frac{\Gamma \rhd \mathcal{D}\left(d_\Gamma, \Gamma' \rhd \mathcal{D}\left(d_{\Gamma'}, \Gamma'' \rhd \mathcal{D}\left(d_{\Gamma''}, \dots\right)\right)\right) \vdash \varepsilon \xrightarrow{reduction} \varepsilon_{nf}}{\Gamma' \rhd \mathcal{D}\left(d_{\Gamma'}, \Gamma'' \rhd \mathcal{D}\left(d_{\Gamma''}, \dots\right)\right) \vdash \varepsilon_{nf} \oplus d_\Gamma} \quad (20)$$

Because of the particular feature associated with the *call-by-name* evaluation strategy, the process $\Gamma' \rhd \mathcal{D}\left(d_i, d_j, \dots, \Gamma'' \rhd \mathcal{D}(d_m, d_n, \dots)\right)$ in formula (20) is not applied until all $\varepsilon$ in $\Gamma$ have been eliminated completely.

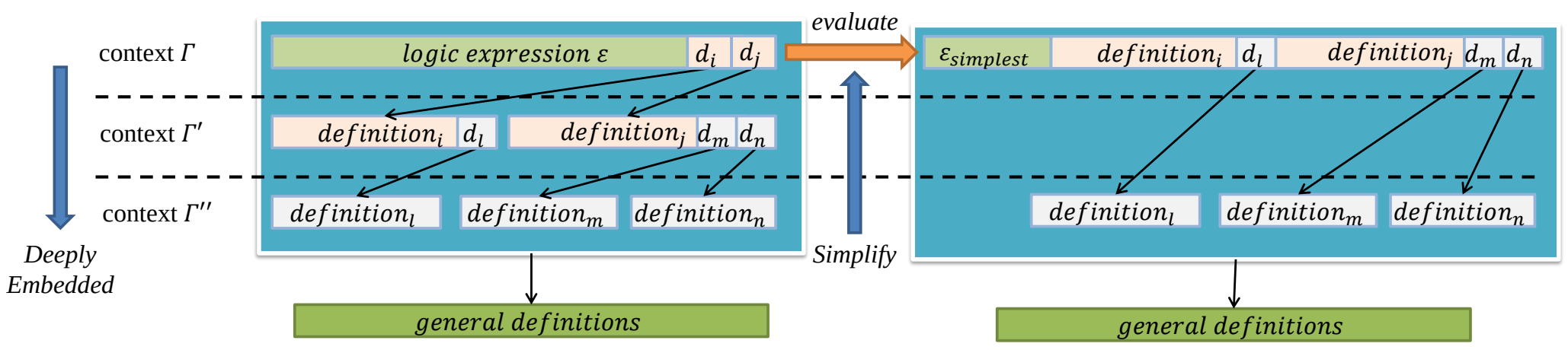

Figure 11. Deeply embedded structure for large functions

In addition, we also note that Solution 2 improves the level of reusability because general definitions, such as address mapping, can be reused by other definitions rather than being redefined. The rules for building definitions can be summarized by the following Principle.

***Principle*** (*Classification*) A pattern matching result should not contain any explicit sub-matching results, and any top-level structure terms should not be transformed as parameters in function calls.

Following *Classification*, the basic expressions $e$ are separately classified into different contexts, and the matching results $c_{sub_i}$ and $c_{nosub_i}$ and other complex terms $r_i$ are separately encapsulated into definitions in different contexts. The effect of this process on $size(\mathcal{FI})$ is illustrated as follows.

$$size(\mathcal{FI}) \equiv \left(\sum_{i=0}^{n}\left(es_i + ds_i + c_{nosub_i} + [r_0 \quad \cdots \quad r_i] * \begin{bmatrix} size_0 \\ \cdots \\ size_i \end{bmatrix}\right) * c_{sub_{i-1}}! + \sum_{i=0}^{n} e_i\right)$$
$$\Rightarrow \left(\left[e_{\Gamma_0} \oplus d_{\Gamma_0}\right] \backsim_{\Gamma_0 \lhd \Gamma_1} \left(\left[e_{\Gamma_1} \oplus d_{\Gamma_1}\right] \backsim_{\Gamma_1 \lhd \Gamma_2} \left(\ldots\left(\left[e_{\Gamma_i} \oplus d_{\Gamma_i}\right] \backsim_{\Gamma_i \lhd \Gamma_{i+1}} (\ldots)\right)\right)\right)\right)$$
$$\Rightarrow \left(\left[e_{\Gamma_0} \oplus d_{\Gamma_0}\right] \backsim_{\Gamma_0 \lhd \overline{\Gamma_d}}\right) \quad (21)$$

Here, we assign $e_{\Gamma_i}$ as the set of basic expressions in the context $\Gamma_i$, and assign $d_{\Gamma_i}$ as the set of all bound names of definitions, which are the entry points of the respective definition bodies. In addition, according to *Classification*, the sub-matching results are defined separately from the super-matching results, and the top-level structure terms are forbidden from being transformed into the sub-definition interface. Therefore, $c_{nosub_i}$ and $c_{sub_i}$ are eliminated from formula (21), and $size(\mathcal{FI})$ is further simplified using $\overline{\Gamma_d}$, which represents the set of deep contexts.

According to the *call-by-name* evaluation strategy, definitions $d_\Gamma$ do not occupy the computing resource. Moreover, the body of the definitions $d_{\Gamma_i}$, which is defined in the $\backsim_{\Gamma_i \lhd \overline{\Gamma_d}}$ process, will not be unfolded until either $e_{\Gamma_i}$ is eliminated as the normal form or the definitions in $d_{\Gamma_i}$ are necessary. This is defined as follows.

$$infor_\Gamma \equiv e_\Gamma \oplus d_\Gamma \oplus size(statement) \quad (22)$$

In this way, only the basic expressions $e_{\Gamma_i}$, the definition entry points $d_{\Gamma_i}$ in context $\Gamma_i$, and the bodies of the definitions in $d_{\Gamma_i}$ will be hidden in deeper contexts, and the trusted core of proof assistants needs only to evaluate $e_{\Gamma_i}$. Accordingly, the value of $infor_{size}$ for $\Gamma_c$ can be simplified as follows.

$$infor_{size} \equiv es_\Gamma + ds_\Gamma + size(statement) \quad (23)$$

Although this process will increase the number of times the *unfold* operation must be conducted, the influence of this increase on the computational efficiency of $\mathcal{FI}$ is negligible because the *unfold* operation is one of the simplest atomic operations. Therefore, the computational load of the *unfold* operation in $\Gamma_c$ on the computing resource is much less than the original evaluation process. In addition, the number of times the *unfold* operation must be conducted is less than or equal to the number of definitions $d_\Gamma$ in $\Gamma_c$, and the number of $e_\Gamma \oplus d_\Gamma$ operations in $\Gamma_c$ for any $\mathcal{FI}$ is practically fixed at a constant value. Thus, the value of $infor_{size}$ for $\Gamma_c$ is influenced only by the complexity of a single statement, which is denoted by $size(statement)$. Moreover, the requirement that $e_\Gamma$ be simplified as the normal form eliminates irrelevant logic expressions from the computational load. Therefore, $infor_{size}$ is significantly reduced, such that the computational load of the trusted core remains within a manageable range. Thus, the IRE problem is solved.

***Solution* 3** (***Concurrent reduction***) Finally, the cause of CR can be easily solved by defining different *pump* limitations for every layer, as shown in Figure 12. Here, we modify the *K*-limitation structure adopted in each layer, respectively, which solves the CR problem completely.

$$K_{stt} \xrightarrow{deconstrcuted} \sigma_0 \Downarrow_{stt} \xrightarrow{unfolded} \mathcal{E}S_{stt} \xRightarrow{simplify} [\![\sigma_0{}']\!]$$

*Statement layer*

$$K_e \xrightarrow{deconstrcuted} \sigma_1 \Downarrow_e \xrightarrow{unfolded} \mathcal{E}S_e \xRightarrow{simplify} [\![\sigma_1{}']\!]$$

*Expression layer*

$$K_v \xrightarrow{deconstrcuted} \sigma_2 \Downarrow_v \xrightarrow{unfolded} \mathcal{E}S_v \xRightarrow{simplify} [\![\sigma_2{}']\!]$$

*Value layer*

Figure 12. New *K*-limitation structure for FEther to alleviate the CR problem

### 5.2 Case Study

We applied the three proposed optimization schemes in the development process of a new version of FEther for FSPVM-E, and employed FSPVM-E to execute (i.e., verify) the very simple code segment in Example 1. Compared with the result given in Figure 3 for the non-optimized version of FEther, the results in Figure 13 indicate that the execution time decreased from 92.546 s to 0.033 s for the optimized version of FEther. As such, the optimized version requires just 3/10000 of the time required by the non-optimized version. The details regarding this new version of FEther will be introduced in a subsequent technological report.

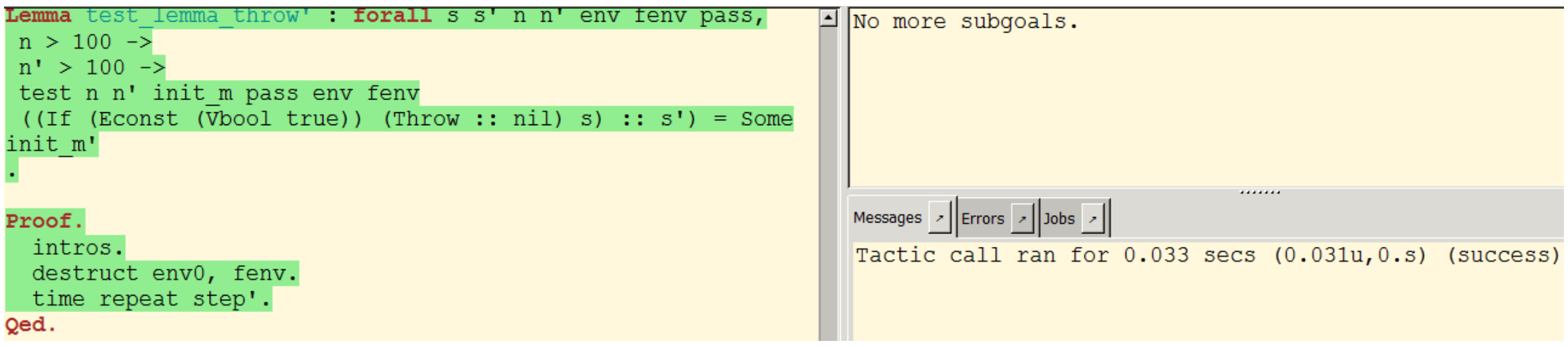

Figure 13. Simple example for testing the execution efficiency of FEther after optimization

In addition, we also tested the optimized version of FEther under an identical experimental environment and with an equivalent data set as those employed for the results obtained in Figure 4 by the non-optimized version of FEther. As shown in Figure 14, the purple line is the peak execution times of *FRWprograms* constructed using specific instructions, and the red line is the peak execution times of *FRWprograms* constructed using abstract instructions. Compared with the results in Figure 4, we note that both program types exhibit a linear increase in execution time with respect to an increasing number of lines, rather than exponentially, as was obtained using the non-optimized version of FEther.

These experimental results verify that the optimization schemes provide results that conform with our analyses of the causes of CBNT, IRE, and CR. Moreover, the experiments certify that these schemes can optimize FEther, and improve the execution efficiency of the proposed FSPVM-E significantly.

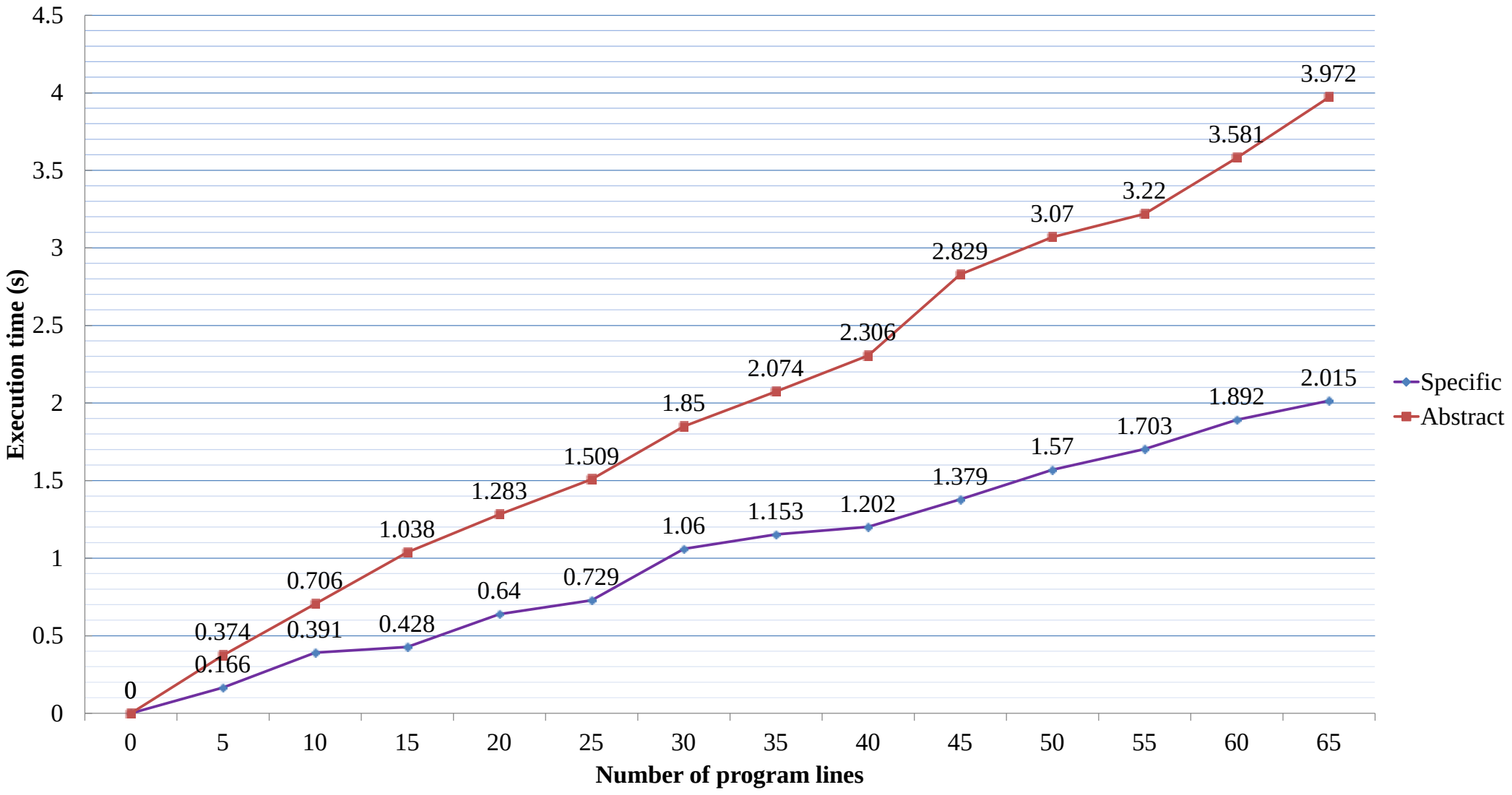

Figure 14. Maximum evaluation times of FEther after optimization for example smart contracts given in [4]

## 6. Conclusions and future work

In this paper, we presented analyses of the issues denoted as *call-by-name termination*, *information redundancy explosion*, and *concurrent reduction* that reduce the evaluation efficiency of formal interpreters adopted in an FSPVM framework and other large programs developed in higher-order theorem proving assistants. We then built abstract models based on these analyses, and developed respective optimization schemes for each issue. Finally, we applied the proposed schemes to optimize the FEther interpreter employed in FSPVM-E. Experimental results verified that the execution efficiency of FEther was improved significantly. We are presently pursuing the optimization and certification of FSPVM-E. Then, we will extend FSPVM-E to support the formal verification of smart contracts on the EOS blockchain platform, which includes the formalization of a subset of the C++ language and the respective interpreter based on the GERM framework.

## Acknowledgement

The authors wish to thank Marisa for the kind assistance.